%% file: preprint.tex
\documentclass[11pt, oneside]{article}   	
\usepackage{geometry}                		
\usepackage{graphicx}
\usepackage{amsmath}
\usepackage{amsfonts}
\usepackage{float}
\usepackage{import}
\usepackage{url}
\usepackage{color}
\usepackage{multirow}
\usepackage{subfigure}
\usepackage{appendix}
\usepackage{pgfplots}
\usepackage{siunitx}
\usepackage{tikz}
\usepackage{tkz-euclide}
\usepackage{stanli}
\usepackage{authblk}

\title{A modified bond model for describing isotropic linear elastic material behaviour with the particle method}
\author{R. G. Venkateswaran\thanks{Correspondence: r.venkateswaran@tu-braunschweig.de}  }
\author{U. Kowalsky}
\author{D. Dinkler}

\affil{Institute of Structural Analysis\\
Technische Universit\"at Braunschweig\\
Braunschweig, Germany\\}
\date{}							

\begin{document}
\maketitle

\begin{abstract}
Particle based methods such as the Discrete Element Method and the Lattice Spring Method may be used for describing the behaviour of isotropic linear elastic materials. However, the common bond models employed to describe the interaction between particles restrict the range of Poisson's ratio that can be represented. In this paper, to overcome the restriction, a modified bond model that includes the coupling of shear strain energy of neighbouring bonds is proposed. The coupling is described by a multi-bond term that enables the model to distinguish between shear deformations and rigid-body rotations. The positive definiteness of the strain energy function of the modified bond model is verified. To validate the model, uniaxial tension, pure shear, pure bending and cantilever bending tests are performed. Comparison of the particle displacements with continuum mechanics solution demonstrates the ability of the model to describe the behaviour of isotropic linear elastic material for values of Poisson's ratio in the range $0 \leq \nu < 0.5$.

\flushleft{\bf Keywords: }DEM, strain energy, multi-bond, shear, rigid-body rotation, Poisson's ratio
\end{abstract}

\section{Introduction}
\label{sec: intro}
Regarding the atomic scale a material is made up of atoms which are known to interact with each other through attractive and repulsive forces. Extrapolating this model to the macroscopic scale is the fundamental idea of modelling continua with interacting discrete particles. Referring to the historical development of the theory of elasticity \cite{Love:2013aa,Capecchi:2010aa} one realises that the mechanicians of the nineteenth century have made substantial contribution to the development of this alternate discrete model of elasticity. Apart from elasticity, it allows one to describe the failure of material as the absence of interaction between two previously interacting material points. Two of the most widely used particle methods are the Lattice Spring Model (LSM) and the discrete element method (DEM) initially developed for modelling the motion of granular media. The LSM influenced by crystal elasticity discretises the domain with material points in an ordered manner, whereby each point interacts with each other by means of an elastic spring. The DEM initially developed by Cundall \cite{Cundall:1979aa} has been extended to model solids such as rocks by the bonded-particle method (BPM) where two particles remain bonded together as long as a critical bond deformation is reached \cite{Potyondy:2004aa}. Although these methods are different in their implementation they are similar in the fact that they use bond models to describe the interaction between two particles. Several bond models exist in the literature and a brief summary of the most commonly used models is provided here. For a detailed review on the various bond models refer to \cite{Ostoja-Starzewski:2002aa}.

The fundamental bond model is a Hookean spring which penalises change from the reference length between two interacting particles. For a solid discretised with an infinite number of material points connected by such springs, the Poisson's ratio is restricted to $1/3$ in the case of plane stress and $1/4$ for plane strain and in three-dimensions \cite{Griffiths:2001aa,Ostoja-Starzewski:2002aa,Zhao:2011aa}. This model has one stiffness parameter to describe macroscopic isotropic elasticity which is on the other hand described by two independent material parameters. Therefore, additional crystal symmetry conditions called \textit{Cauchy symmetry} are fulfilled and lead to the restriction on Poisson's ratio \cite{Love:2013aa,Capecchi:2010aa}. In order to overcome the restriction, Born \cite{Born:1954aa} introduced a shear spring in addition to the longitudinal spring. The shear springs can be understood as a penalisation of the change in orientation from the reference configuration. This model has two stiffness parameters available for calibration. However, for an infinite number of such bonds, the Poisson's ratio is restricted to the range $0 \le \nu \le \frac{1}{3}$ and $0 \le \nu \le \frac{1}{4}$ for the case of plane stress and plane strain respectively. Apart from these restrictions of Poisson's ratio, the model cannot distinguish between rigid-body rotation and shear deformation \cite{Hassold:1989aa,Buxton:2001aa,Keating:1966aa}. Instead of using shear springs, the Kirkwood-Keating spring model \cite{Kirkwood:1939aa,Keating:1966aa} uses an angular spring to penalise the angular motion. However, this model is known to be nonlinear due to the angular terms and it offers no substantial advantage in comparison to the \textit{Born model} \cite{Buxton:2001aa}. The Lattice Beam model (LBM) uses a Hookean spring to allow longitudinal forces and a beam to allow shear force and bending moment and is widely used in DEM. However, in DEM the pure bending modes are neglected \cite{Smilauer:2015aa,Griffiths:2001aa}. The beam model has the similar restriction on the Poisson's ratio as the \textit{Born model} but is capable of distinguishing between rigid-body rotation and shear deformation due to inclusion of the rotational degree of freedom.

In order to overcome the limited range of Poisson's ratio that can be represented by these models, Zhao \cite{Zhao:2011aa} evaluated the shear strains in the \textit{Born model} using a particle strain tensor instead of particle displacements. This tensor is an approximate measure of the continuum strain tensor and is obtained by the least square method using the information of displacements of the particle under consideration and its neighbouring particles. The calculation of the particle strain tensor is computationally expensive and for a practical simulation with random particle arrangements there exists ambiguity regarding its existence. In such scenario only Hookean springs are used. Celigueta \cite{Celigueta:2017aa} proposed a nonlocal contact law for the DEM, where apart from the overlap of two interacting particles in contact also the influence of forces in the surrounding of this contact is included. By means of this term a better description of continuum elasticity was obtained in comparison to the classical bond model of DEM. However, in order to calculate this term information regarding the contact area between two particles in contact and their surroundings are required. Additionally, the nonlocal stress tensor of each particle and its neighbours are required which results in a computationally expensive method.

In this work, a modification of the \textit{Born model} is proposed by introducing a multi-bond strain energy term that is capable of distinguishing between shear deformation and rigid-body rotation for the case of two-dimensional elasticity. Two new multi-bonds called \textit{L-bond} and \textit{X-bond} employing the proposed coupling of shear strain energy are introduced. Based on positive definiteness of the strain energy function, we also interpret that the restriction of the \textit{Born model} up to a certain Poisson's ratio results from incapability to distinguish between shear deformation and rigid-body rotation. With the modified model it is shown that the strain energy function remains positive definite for values of Poisson's ratio in the range $0 \leq \nu < 0.5$. To validate the modified model, the results of a thin plate under uniaxial tension, pure shear, pure bending and cantilever bending are compared with the respective plane stress continuum mechanics solution.

\section{Unit-cells with \textit{Born model}}
\label{sec: unit-cell with Born bond model}
\begin{figure}
\centering
\subfigure[]{\input{tikz/figure2a}}
\hspace{5mm}
\subfigure[]{\input{tikz/figure2b}}
\caption{(a) Springs used in a bond (b) Coordinate system of a generic bond}
\label{fig: generic bond}
\end{figure}
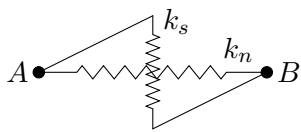
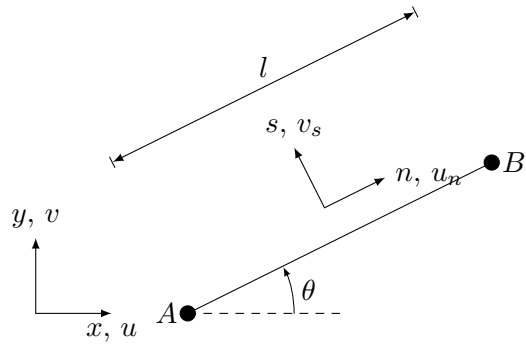

An arbitrary rectangular domain discretised with rigid circular particles of radius $r$ is shown in Figure \ref{fig: domain discretised with chosen unit-cell}(a). Each particle is bonded with its immediate neighbours (separated by $2r$) and also with its second neighbours (separated by $2 \sqrt{2} r$). Similar to the works \cite{Griffiths:2001aa,Ockelmann:2017aa}, the fundamental building block called as a \textit{unit-cell} is extracted from the discretised domain and is shown in Figure \ref{fig: domain discretised with chosen unit-cell}(b). This square unit-cell is the combination of two unit-cells: one which contains only the bonds with the immediate neighbours and another that contains only the bonds between second neighbours. They are called first-neighbour and second-neighbour unit-cells as shown in Figure \ref{fig: domain discretised with chosen unit-cell}(c) and \ref{fig: domain discretised with chosen unit-cell}(d) respectively. Although other configurations such as a triangle or a hexagon can be used as a unit-cell, the square configuration is used here due to its simplicity. The modification proposed in this paper is not restricted to a square configuration and it can be used also regarding other regular configurations. The \textit{Born model} which is applied here employs springs in normal direction and tangential direction with stiffness parameters $k_{n}$ and $k_{s}$ as shown in Figure \ref{fig: generic bond}(a). A generic bond oriented at angle $\theta$ to the global coordinate system along with its local and global displacement components is shown in Figure \ref{fig: generic bond}(b).

\begin{figure}
\centering
\subfigure[]{\input{tikz/figure1a.tex}}
\hspace{1cm}
\subfigure[]{\input{tikz/figure1b.tex}}
\subfigure[]{\input{tikz/figure1c.tex}}
\subfigure[]{\input{tikz/figure1d.tex}}
\caption{(a) Domain discretised with the chosen unit-cell (b) Square unit-cell obtained from the combination of (c) First neighbour unit-cell and (d) Second neighbour unit-cell}
\label{fig: domain discretised with chosen unit-cell}
\end{figure}
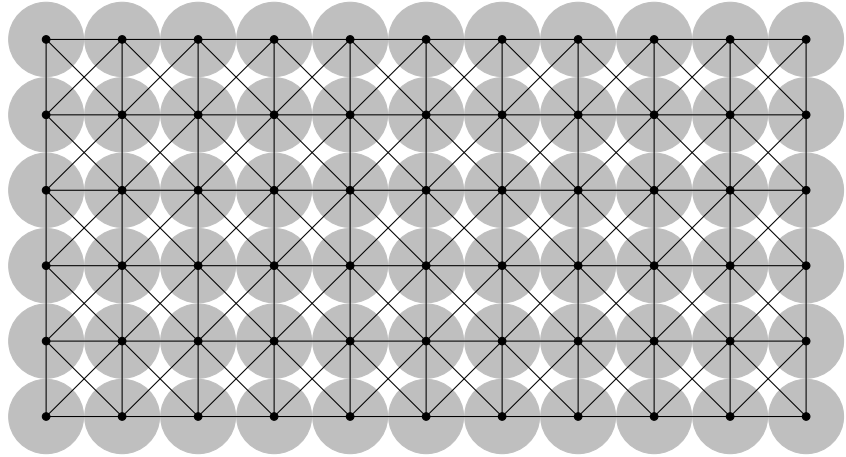
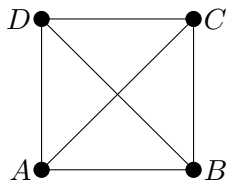
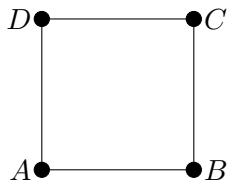
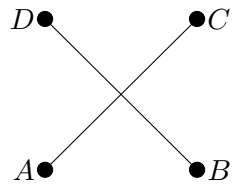

Regarding Figure \ref{fig: generic bond}(a), the strain energy stored in the bond $\Pi_{b}$ is given in terms of quantities with respect to the bond coordinate system as
\begin{align}
	\Pi_{b} &= \frac{1}{2} \, k_{n} (u_{B}^{n} - u_{A}^{n})^{2} + \frac{1}{2} \, k_{s} (v_{B}^{s} - v_{A}^{s})^{2} \label{eqn: strain energy in bond in terms of local displacements}\\
	&= \frac{l^{2}}{2} \, \big( k_{n} \, (\epsilon_{nn})^{2} + k_{s} \, (\epsilon_{sn})^{2} \big), \label{eqn: strain energy in bond in terms of local strains}
\end{align}
where $\epsilon_{nn} = \frac{u_{B}^{n} - u_{A}^{n}}{l}$ and $\epsilon_{sn} = \frac{v_{B}^{s} - v_{A}^{s}}{l}$ are the local longitudinal and shear strain respectively and $l$ is the length of the undeformed bond.

After substituting the local strain quantities in terms of the components of the global strain tensor by the transformation relations provided by Griffiths \cite{Griffiths:2001aa} the strain energy in a bond is obtained as 
\begin{align}
\label{eqn: strain energy in bond in terms of global strains}
\Pi_{b} = \frac{l^{2}}{2} \, \bigg( k_{n} \, (c^{2} \epsilon_{xx} + s c (\epsilon_{xy} + \epsilon_{yx}) + s^{2} \epsilon_{yy})^{2} + k_{s} \, (c^{2} \epsilon_{yx} + s c (\epsilon_{yy} - \epsilon_{xx}) - s^{2} \epsilon_{xy})^{2} \bigg)
\end{align}
where, $c = \cos\theta$ and $s = \sin\theta$. By substituting the orientation $\theta$ of the individual constituent bonds in Equation (\ref{eqn: strain energy in bond in terms of global strains}), the strain energy in the first and second neighbour unit-cells are obtained as
\begin{align}
\label{eqn: strain energy in first neighbour unit-cell}
\begin{aligned}
	\Pi_{uc_{1}} &= \Pi_{uc_{1}}^{normal} + \Pi_{uc_{1}}^{shear} \\
	&= \frac{k_{n_{1}}^{b} \, l^{2}}{4}  \bigg[ (\epsilon_{xx})^{2} + (\epsilon_{yy})^{2} + (\epsilon_{xx})^{2} + (\epsilon_{yy})^{2} \bigg] \\
	&+ \frac{k_{s_{1}}^{b} \, l^{2}}{4}  \bigg[ (\epsilon_{yx})^{2} + (-\epsilon_{xy})^{2} + (\epsilon_{yx})^{2} + (-\epsilon_{xy})^{2} \bigg],
\end{aligned}
\end{align}
\begin{align}
\label{eqn: strain energy in second neighbour unit-cell}
\begin{aligned}
	\Pi_{uc_{2}} &= \Pi_{uc_{2}}^{normal} + \Pi_{uc_{2}}^{shear} \\
	&= k_{n_{2}}^{b} \, l^{2} \bigg[ \bigg(\frac{\epsilon_{xx}}{2} + \frac{\epsilon_{xy}}{2} + \frac{\epsilon_{yx}}{2} + \frac{\epsilon_{yy}}{2}\bigg)^{2} + \bigg(\frac{\epsilon_{xx}}{2} - \frac{\epsilon_{xy}}{2} - \frac{\epsilon_{yx}}{2} + \frac{\epsilon_{yy}}{2}\bigg)^{2} \bigg] \\
	&+ \frac{k_{s_{1}}^{b} \, (\sqrt{2}l)^{2}}{2} \bigg[ \bigg(\frac{\epsilon_{yx}}{2} - \frac{\epsilon_{xy}}{2} - \frac{\epsilon_{xx}}{2} + \frac{\epsilon_{yy}}{2}\bigg)^{2} + \bigg(\frac{\epsilon_{xx}}{2} - \frac{\epsilon_{xy}}{2} + \frac{\epsilon_{yx}}{2} - \frac{\epsilon_{yy}}{2}\bigg)^{2} \bigg],
\end{aligned}
\end{align}
where $k_{n_{1}}^{b}$,  $k_{s_{1}}^{b}$ are the normal and shear stiffness of first neighbour bonds and similarly $k_{n_{2}}^{b}$, $k_{s_{2}}^{b}$ are the stiffness parameters of second neighbour bonds. All bonds in the same shear plane were assigned the same stiffness $k_{s_{1}}^{b}$. Every bond in the first neighbour unit-cell contributes only $1/2$ to the strain energy of the unit-cell due to periodicity (every first neighbour bond is shared by two unit-cells). However the bonds in the second neighbour unit-cell belong completely to a single unit-cell. The strain energy density of the square unit-cell is given by
\begin{align} \label{eqn: strain energy density in unit-cell}
	\Pi_{uc} &= \Pi_{uc_{1}} + \Pi_{uc_{2}} \nonumber \\
	e_{uc} &= \frac{\Pi_{uc}}{l^{2}t},
\end{align}
where $t$ is the thickness of the domain. 

\begin{figure}
\centering
\subfigure[]{\input{tikz/figure3a}} \hspace{1cm}
\subfigure[]{\input{tikz/figure3b}}
\subfigure[]{\input{tikz/figure3c}} \hspace{1cm}
\subfigure[]{\input{tikz/figure3d}}
\caption{(a) Split up of first neighbour unit-cell bonds (b) Assembly of first neighbour bonds (c) Split up of second neighbour unit-cell bonds (d) Assembly of second neighbour bonds}
\label{fig: reason for rotational variance square uc}
\end{figure}
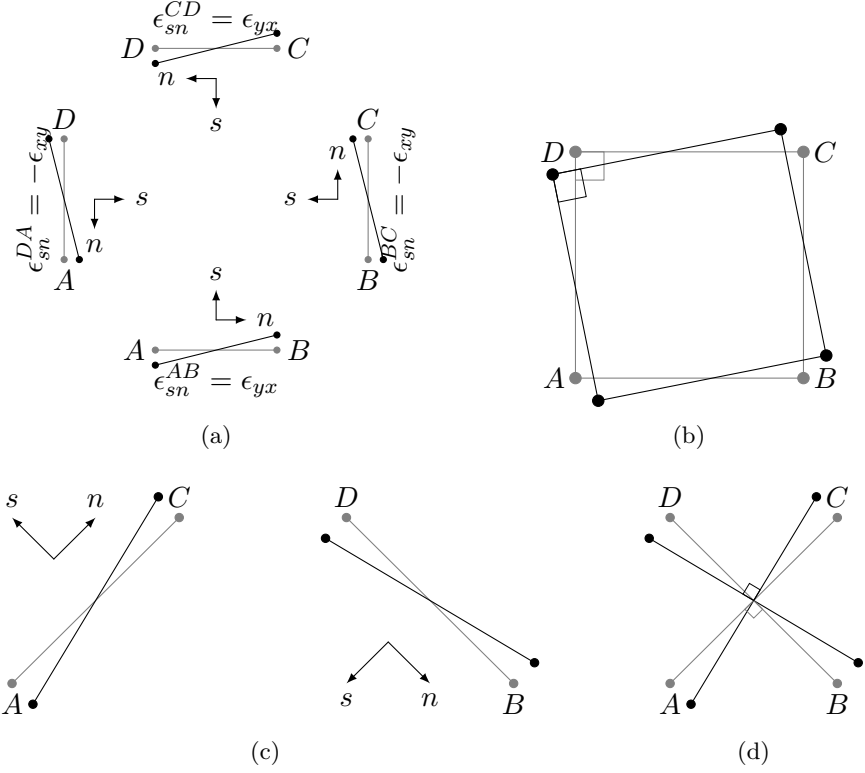

As described before, the \textit{Born model} cannot distinguish between rigid-body rotation and shear \cite{Keating:1966aa,Hassold:1989aa,Zhao:2011aa,Buxton:2001aa}. In Figure \ref{fig: reason for rotational variance square uc}(a) an exploded view of each constituent bond of the unit-cell with first neighbour bonds is shown along with its local and global shear strains. The bonds are assembled back together in Figure \ref{fig: reason for rotational variance square uc}(b) and one observes that although the final geometry describes pure rotation, the strain energy is not zero. Thus strain energy is stored in the unit-cell due to shear of individual bonds although all angles subtended between neighbour bonds are $90^{\circ}$. This can also be observed in Equation (\ref{eqn: strain energy in first neighbour unit-cell}) where the contribution of shear strain energy $\Pi_{uc_{1}}^{shear}$ remains non-zero for any combination of $\epsilon_{xy}$ and $\epsilon_{yx}$ except for the trivial case where $\epsilon_{xy} = \epsilon_{yx} = 0$. This interpretation holds also for the unit-cell with the second neighbour bonds as shown in Figure \ref{fig: reason for rotational variance square uc}(c), (d).

\section{Modified bond model}
\label{sec: improved bond model}
To overcome the limitations explained in Section \ref{sec: unit-cell with Born bond model}, a modified model is proposed where the shear strains of two neighbour bonds (multi-bond) are coupled. With this coupling, strain energy is stored only in the case of shear and not in the case of rotation. Here again, the unit-cell made up of only the first neighbour bonds is looked at first and the unit-cell made up of the second neighbour bonds afterwards.

\subsection{\textit{L-bond}} 
\label{subsec: L-bond}
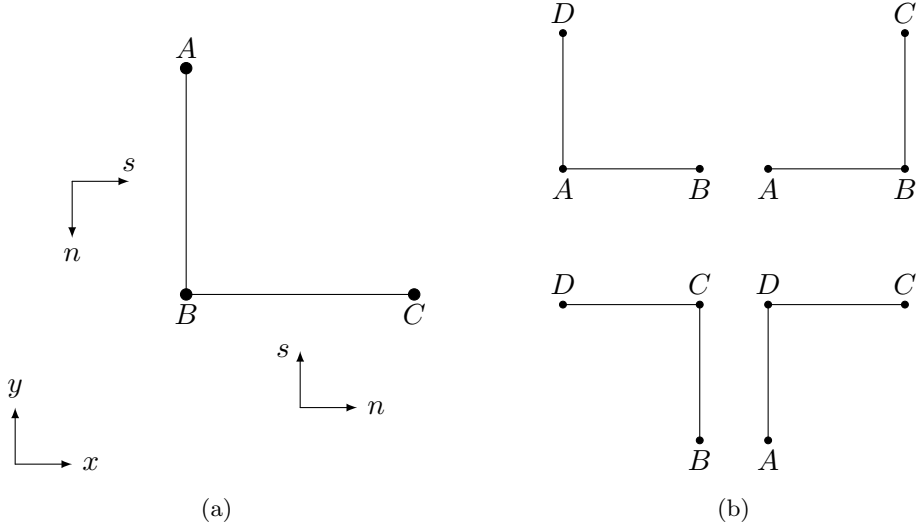
\begin{figure}[H]
\centering
\subfigure[]{\input{tikz/figure4a}} \hspace{1cm}
\subfigure[]{\input{tikz/figure4b}}
\caption{(a) Generic \textit{L-bond} (b) First neighbour unit-cell made up with \textit{L-bonds}}
\label{fig: first uc with l-bonds}
\end{figure}

For a generic \textit{L-bond} ABC made up of two first neighbour bonds AB and BC as shown in Figure \ref{fig: first uc with l-bonds}(a), the strain energy due to normal strain remains unchanged as before. However the shear strains are combined in such a way that strain energy is stored only for shear deformation and not for rigid-body rotation. The strain energy stored in a generic \textit{L-bond} with length $l_{AB} = l_{BC} = l$ is given by
\begin{align} \label{eqn: strain energy generic lbond in local strain components}
	\Pi_{ABC} = \frac{1}{2} k_{n_{1}}^{m} \bigg[ \big( \epsilon_{nn}^{AB} \, l \big)^{2} + \big( \epsilon_{nn}^{BC} \, l \big)^{2} \bigg] + \frac{1}{2} k_{s_{1}}^{m} \bigg[ -\epsilon_{sn}^{AB} \, l + \epsilon_{sn}^{BC} \, l \bigg]^{2},
\end{align}
where $k_{n_{1}}^{m}$ and $k_{s_{1}}^{m}$ are the normal stiffness and shear stiffness parameters employing the modified model respectively. The first neighbour unit-cell is now made up of four L-bonds as shown in Figure \ref{fig: first uc with l-bonds}(b). 

As explained in Section \ref{sec: unit-cell with Born bond model}, a first neighbour bond has a contribution factor of $1/2$. Every bond in the unit-cell is obtained by the combination of two L-bonds. For example, the Bond $AB$ is obtained from $DAB$ and $ABC$. Therefore, in an L-bond the normal stiffness $k_{n_{1}}^{m}$ and the shear stiffness $k_{s_{1}}^{m}$ have a contribution factor of $1/4$. With the transformation relations provided by Griffiths \cite{Griffiths:2001aa}, the local strain components are written in terms of the global strain components and the strain energy of a generic \textit{L-bond} now yields
\begin{align} \label{eqn: strain energy generic lbond in global strain components}
\begin{aligned}
	\Pi_{ABC} = \frac{k_{n_{1}}^{m} \, l^{2}}{8} \, \bigg[ \bigg( \epsilon_{xx} \, c_{AB}^{2} + \big( \epsilon_{xy} + \epsilon_{yx} \big) \, c_{AB} \, s_{AB} + \epsilon_{yy} \, s_{AB}^{2} \bigg)^{2} \\ + \bigg( \epsilon_{xx} \, c_{BC}^{2} + \big( \epsilon_{xy} + \epsilon_{yx} \big) \, c_{BC} \, s_{BC} + \epsilon_{yy} \, s_{BC}^{2} \bigg)^{2} \, \bigg]\\
	+ \frac{k_{s_{1}}^{m} \, l^{2}}{8} \, \bigg[ - \bigg( \epsilon_{yx} \, c_{AB}^{2} + \big( \epsilon_{yy} - \epsilon_{xx} \big) \, c_{AB} \, s_{AB} - \epsilon_{xy} \, s_{AB}^{2} \bigg)\\ + \bigg( \epsilon_{yx} \, c_{BC}^{2} + \big( \epsilon_{yy} - \epsilon_{xx} \big) \, c_{BC} \, s_{BC} - \epsilon_{xy} \, s_{BC}^{2} \bigg) \bigg]^{2},
\end{aligned}
\end{align}
where, $c_{AB} = \cos \theta_{AB}$, $s_{AB} = \sin \theta_{AB}$, $c_{BC} = \cos \theta_{BC}$ and $s_{BC} = \sin \theta_{BC}$. Employing the modified bond model, the strain energy stored in the first neighbour unit-cell is obtained from the sum of strain energy stored in individual \textit{L-bonds}
\begin{align}
	\Pi_{uc_{1}}^{mod} = \Pi_{ABC} + \Pi_{BCD} + \Pi_{CDA} + \Pi_{DAB}.
\end{align}

The strain energy stored in the first neighbour unit-cell employing the modified bond model can be evaluated with respect to the global coordinate system to
\begin{align}
\begin{aligned} \label{eqn: strain energy in first neighbour unit-cell with improved bond model}
\Pi_{uc_{1}}^{mod} &= \frac{k_{n_{1}}^{m} \, l^{2}}{2} \, \big( \epsilon_{xx}^{2} + \epsilon_{yy}^{2} \big) + \frac{k_{s_{1}}^{m} \, l^{2}}{2} \, \big( \epsilon_{xy} + \epsilon_{yx} \big)^{2}\\
	&= \frac{k_{n_{1}}^{m} \, l^{2}}{2} \, \big( \epsilon_{xx}^{2} + \epsilon_{yy}^{2} \big) + \frac{k_{s_{1}}^{m} \, l^{2}}{2} \, \big( \epsilon_{xy}^{2} + \epsilon_{yx}^{2} \big) +  k_{s_{1}}^{m} \, l^{2} \, \epsilon_{xy} \, \epsilon_{yx} \\
	&= \Pi_{uc_{1}}^{normal} + \Pi_{uc_{1}}^{shear} + k_{s_{1}}^{m} \, l^{2} \, \epsilon_{xy} \, \epsilon_{yx}.
\end{aligned}
\end{align}
Comparing the formulation to that of the first neighbour unit-cell employing the \textit{Born model}, see Equation (\ref{eqn: strain energy in first neighbour unit-cell}), an extra term is added due to the coupling of the shear strain energies of the neighbours. With this term the modified model is able to distinguish between shear (change in the angle subtended) and rigid-body rotation (no change in the angle subtended), which can be interpreted as a modification to the \textit{Born model}.

\subsection{\textit{X-bond}}
\label{subsec: X-bonds}
Similar to \textit{L-bonds}, the unit-cell made up of the second neighbour bonds is to be modified such that strain energy is stored only in the case of shear deformations and not for pure rotation. This is achieved with the \textit{X-bonds}. Again the shear strains of two neighbouring bonds are coupled and the normal strains of the bonds remain unchanged. The modified strain energy of the unit-cell made up of one \textit{X-bond} is
\begin{align}
	\Pi_{uc_{2}}^{mod} &= \Pi_{AC} + \Pi_{BD} \nonumber \\
	&= \frac{k_{n_{2}}^{m} \, (\sqrt{2} \, l)^{2}}{2} \bigg[ \big( \epsilon_{nn}^{AC} \big)^{2} + \big( \epsilon_{nn}^{BC} \big)^{2} \bigg] + \frac{k_{s_{2}}^{m} \, (\sqrt{2} \, l)^{2}}{2} \bigg[ \epsilon_{sn}^{AC} - \epsilon_{sn}^{BD}\bigg]^{2},
\end{align}
where, $k_{n_{2}}^{m}$ and $k_{s_{2}}^{m}$ are the normal and shear stiffness parameters employing the modified model respectively. The shear stiffness parameters in the same shear plane are assumed to have the same stiffness parameter $k_{s_{1}}^{m}$. The strain energy stored in the second neighbour unit-cell with respect to the global coordinate system is
\begin{align} 
\begin{aligned} \label{eqn: strain energy in second neighbour unit-cell with improved bond model}
	\Pi_{uc_{2}}^{mod} &= \frac{k_{n_{2}}^{m} \, (\sqrt{2}l)^{2}}{2} \bigg[ \bigg( \frac{\epsilon_{xx}}{2} + \frac{\epsilon_{xy}}{2} + \frac{\epsilon_{yx}}{2} + \frac{\epsilon_{yy}}{2} \bigg)^{2} + \bigg( \frac{\epsilon_{xx}}{2} - \frac{\epsilon_{xy}}{2} - \frac{\epsilon_{yx}}{2} + \frac{\epsilon_{yy}}{2}\bigg)^{2} \bigg]\\
	&+ \frac{k_{s_{1}}^{m} \, (\sqrt{2}l)^{2}}{2} \bigg[ \bigg(\frac{\epsilon_{yx}}{2} - \frac{\epsilon_{xy}}{2} - \frac{\epsilon_{xx}}{2} + \frac{\epsilon_{yy}}{2}\bigg) - \bigg(\frac{\epsilon_{xx}}{2} - \frac{\epsilon_{xy}}{2} + \frac{\epsilon_{yx}}{2} - \frac{\epsilon_{yy}}{2}\bigg) \bigg]^{2}.
\end{aligned}
\end{align}

Upon expansion of Equation (\ref{eqn: strain energy in second neighbour unit-cell with improved bond model}) and comparison with the strain energy of the second neighbour unit-cell, see Equation (\ref{eqn: strain energy in second neighbour unit-cell}), one observes that similar to the first neighbour unit-cell with \textit{L-bonds}, the coupling of shear strain energies results in an extra multi-bond term to distinguish between shear and rotation.

\subsection{Unit-cell with the modified model}
\label{subsec: unit-cell with improved bond model}
The strain energy stored in a square unit-cell is obtained by summing up the strain energy of an unit-cell made with \textit{L-bonds}, Equation (\ref{eqn: strain energy in first neighbour unit-cell with improved bond model}), and of an unit-cell with \textit{X-bonds}, Equation (\ref{eqn: strain energy in second neighbour unit-cell with improved bond model})
\begin{align} \label{eqn: strain energy in unit-cell with improved bond model}
	\Pi_{uc}^{mod} = \Pi_{uc_{1}}^{mod} + \Pi_{uc_{2}}^{mod}.
\end{align}
The strain energy density $e_{uc}^{mod}$ is obtained similar to that of the unit-cell with the \textit{Born model} given in Equation (\ref{eqn: strain energy density in unit-cell}).

\section{Comparison of the modified model and the \textit{Born model}}
\label{sec: comparison of the improved bond model with that of the Born bond model}
In this section, the stiffness parameters of a square unit-cell employing the modified model and the \textit{Born model} are calibrated to the macroscopic elastic material parameters. The calibrated parameters are verified if they satisfy the condition of isotropy. The strain energy function of the unit-cell employing these bond models is verified for its positive definiteness.

\subsection{Calibration of stiffness parameters}
\label{subsec: calibration of stiffness parameters}
\begin{figure}
\centering
\subfigure[\textit{Born} and modified model]{\includegraphics[width=0.49\textwidth]{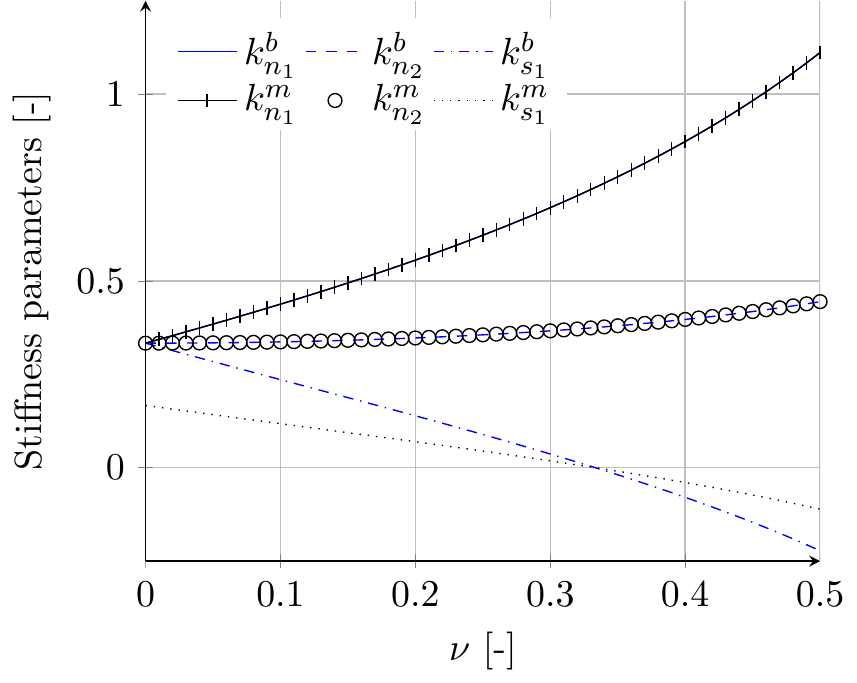}}
\hfill
\subfigure[\textit{Born} and modified model]{\includegraphics[width=0.49\textwidth]{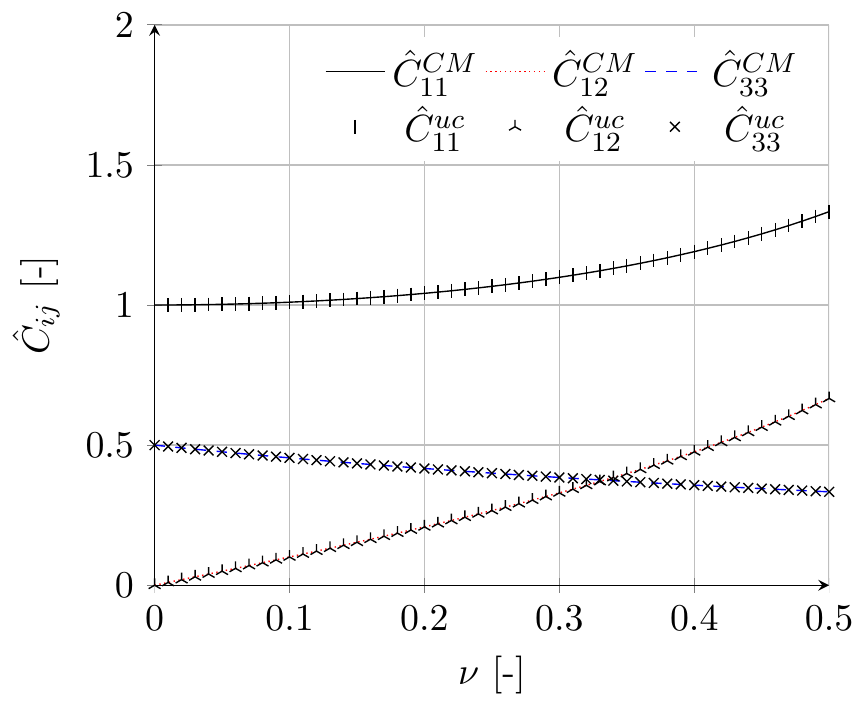}}
\caption{Variation of the stiffness parameters (a) and two-dimensional elasticity tensor components (b) as a function of $\nu$ for plane stress}
\label{fig: plot of calibrated stiffness parameters and elasticity tensor components}
\end{figure}

In order to model the behaviour of an isotropic elastic material, the stiffness parameters of the unit-cell employing the modified model must be calibrated with respect to the macroscopic elastic material parameters. In the literature there exist broadly two approaches:
\begin{enumerate}
	\item A numerical calibration where the stiffness parameters are iteratively calibrated by solving an inverse problem to match the slope of the stress-strain curve of a material in the linear region under uniaxial loading. For more details refer \cite{Zhao:2007aa,Flack:2020aa}. 
	\item An analytical calibration based on the equivalence of strain energy density of the unit-cell $e_{uc}$ with that of an equivalent elastic continuum $e_{co} = \frac{1}{2} \boldsymbol{\sigma} \colon \boldsymbol{\epsilon}$. In particular, the components of the elastic tensor $\mathbb{C}_{uc}$ are derived from the strain energy density of the unit-cell and are compared to the continuum description. 
\end{enumerate}
In this work, the analytical calibration approach is used due to its independence of the fineness of discretisation. The components of the tensor of elasticity of the unit-cell are obtained by differentiating the strain energy density twice with respect to the corresponding strain components. The individual components are summarised in the tensor of elasticity
\begin{align} \label{eqn: components of elasticity tensor of unit-cell}
\begin{aligned}
\mathbb{C}_{uc}^{mod} &=
\begin{bmatrix}
	\hat{C}_{1} & \hat{C}_{2} & 0\\
	\hat{C}_{2} & \hat{C}_{1} & 0\\
	0 & 0 & \hat{C}_{3}
\end{bmatrix},
\end{aligned}
\end{align}
where, $\hat{C}_{1} = k_{n_{1}}^{m} + 2 k_{s_{1}}^{m} + k_{n_{2}}^{m}$, $\hat{C}_{2} = k_{n_{2}}^{m} - 2 k_{s_{1}}^{m}$ and $\hat{C}_{3} = k_{n_{2}}^{m} + k_{s_{1}}^{m}$ are defined for the modified model and $\hat{C}_{1} = k_{n_{1}}^{b} + k_{s_{1}}^{b} + k_{n_{2}}^{b}$, $\hat{C}_{2} = k_{n_{2}}^{b} - k_{s_{1}}^{b}$ and $\hat{C}_{3} = k_{n_{2}}^{b} + \frac{1}{2}\, k_{s_{1}}^{b}$ for the \textit{Born model}. The elasticity tensor of the unit-cell has three components similar to that of a planar continuum. By comparing the three components $\hat{C}_{1}$, $\hat{C}_{2}$ and $\hat{C}_{3}$ with those of the planar continuum elasticity tensor, the stiffness parameters are calibrated to the macroscopic Young's modulus $E$ and the Poisson's ratio $\nu$. The calibrated stiffness parameters for the case of plane stress and plane strain are summarised in Table \ref{table: calibrated stiffness parameters of unmodified and improved square unit-cell}.
\begin{table}[H]
\centering
\caption{Calibrated stiffness parameters of square unit-cell}
\label{table: calibrated stiffness parameters of unmodified and improved square unit-cell}
\normalsize
\begin{tabular}[]{c c c c c}\hline
\noalign{\smallskip}Bond model & Type & \multicolumn{3}{c}{Stiffness parameters}\\ \noalign{\smallskip}\hline
\noalign{\smallskip}\phantom{Bond model} & \phantom{Type} & $k_{n_{1}}^{b}$ & $k_{s_{1}}^{b}$ & $k_{n_{2}}^{b}$\\ \noalign{\smallskip}\cline{3-5}
\noalign{\bigskip}\multirow{2}{*}{\textit{Born}} & Plane stress & $\frac{E \, t \, (1 + 3\nu)}{3 \, (1 + \nu) \, (1 - \nu)}$ & $\frac{E \, t \, (1 - 3\nu)}{3 \, (1 + \nu) \, (1 - \nu)}$ & $\frac{E \, t}{3 \, (1 + \nu) \, (1 - \nu)}$\\ \noalign{\bigskip}
& Plane strain & $\frac{E \, t \, (1 + 2\nu)}{3 \, (1 + \nu) \, (1 - 2\nu)}$ & $\frac{E \, t \, (1 - 4\nu)}{3 \, (1 + \nu) \, (1 - 2\nu)}$ &  $\frac{E \, t \, (1 - \nu)}{3 \, (1 + \nu) \, (1 - 2\nu)}$ \\ \noalign{\bigskip}\hline
\noalign{\smallskip}\phantom{Bond model} & \phantom{Type} & $k_{n_{1}}^{m}$ & $k_{s_{1}}^{m}$ & $k_{n_{2}}^{m}$\\ \noalign{\smallskip}\cline{3-5}
\noalign{\bigskip}\multirow{2}{*}{Modified} & Plane stress & $\frac{E \, t \, (1 + 3\nu)}{3 \, (1 + \nu) \, (1 - \nu)}$ & $\frac{E \, t \, (1 - 3\nu)}{6 \, (1 + \nu) \, (1 - \nu)}$ & $\frac{E \, t}{3 \, (1 + \nu) \, (1 - \nu)}$\\ \noalign{\bigskip}
& Plane strain & $\frac{E \, t \, (1 + 2\nu)}{3 \, (1 + \nu) \, (1 - 2\nu)}$ & $\frac{E \, t \, (1 - 4\nu)}{6 \, (1 + \nu) \, (1 - 2\nu)}$ & $\frac{E \, t \, (1 - \nu)}{3 \, (1 + \nu) \, (1 - 2\nu)}$\\ \noalign{\bigskip} \hline
\end{tabular}
\end{table} 
As expected for a \textit{Born model} there exists a restriction on the Poisson's ratio at $\nu = 1/4$ for plane strain and $\nu = 1/3$ for plane stress. For higher values negative shear stiffness is obtained. This holds true for the modified model as well. The normal stiffness parameters of the unit-cell employing the \textit{Born model} and the modified model are identical since only the shear strain energy components were coupled in the modified bond model. The stiffness parameters and the elasticity constants of the unit-cell are plotted as a function of the Poisson's ratio $\nu$ for both bond models in Figure \ref{fig: plot of calibrated stiffness parameters and elasticity tensor components} for the case of plane stress. The stiffness parameters are normalised with respect to the Young's modulus $E$ and the thickness of the specimen $t$ to obtain a qualitative behaviour.

From the Figure \ref{fig: plot of calibrated stiffness parameters and elasticity tensor components}(b) we observe that the components of the elasticity tensor of the unit-cell agree exactly with the equivalent continuum components in the range $0 \le \nu \le 0.5$. In order to ensure the equality of the elasticity constants, the stiffness parameters of the unit-cell must take on the values as shown in Figure \ref{fig: plot of calibrated stiffness parameters and elasticity tensor components}(a). Therefore, we interpret the negative value of shear stiffnesses ($k_{s_{1}}^{b}$, $k_{s_{1}}^{m}$) for $\nu > 1/3$ as a necessity to ensure this equality. Although a negative stiffness parameter of an individual spring seems unintuitive, in the case of a response of the unit-cell the elasticity tensor components are crucial.

\subsection{Check for isotropy}
\label{subsec: check for isotropy}
Apart from calibrating the stiffness parameters to the macroscopic isotropic elastic material parameters it must also be checked whether and in which case they satisfy the condition for macroscopic isotropy. A material is considered to exhibit isotropy if the anisotropy factor $\Lambda = 1$. The anisotropic factor for the chosen unit-cell is calculated with the components of the elasticity tensor for both bond models as
\begin{align}
\text{\textit{Born model}:} 
\hspace{5mm}
\Lambda &= \frac{2 \, \hat{C}_{3}}{\hat{C}_{1} - \hat{C}_{2}} = \frac{2 k_{n_{2}}^{b} + k_{s_{1}}^{b}}{\phantom{2 \, }k_{n_{1}}^{b} + 2 k_{s_{1}}^{b}} \, %
\begin{cases}
	= 1, & 2 k_{n_{2}}^{b} = k_{n_{1}}^{b} + k_{s_{1}}^{b}\\
	\neq 1, & \text{otherwise}
\end{cases}\\
\text{Modified model:} 
\hspace{5mm}
\Lambda &= \frac{2 \, \hat{C}_{3}}{\hat{C}_{1} - \hat{C}_{2}} = \frac{2 k_{n_{2}}^{m} + 2 k_{s_{1}}^{m}}{\phantom{2 \, }k_{n_{1}}^{m} + 4 k_{s_{1}}^{m}} \, %
\begin{cases}
	= 1, & 2 k_{n_{2}}^{m} = k_{n_{1}}^{m} + 2 k_{s_{1}}^{m}\\
	\neq 1, & \text{otherwise}.
\end{cases}
\end{align}
This shows that for isotropic elasticity ($\Lambda = 1$) the unit-cell has only two independent stiffness parameters and a condition that ensures isotropy for all values of $E$ and $\nu$. By substituting the stiffness parameters from Table \ref{table: calibrated stiffness parameters of unmodified and improved square unit-cell} in the condition for isotropy it can be shown that this condition holds for both plane stress and plane strain.

\subsection{Positive definiteness of strain energy}
\label{subsec: positive definiteness of strain energy}
The square unit-cell shown in Figure \ref{fig: domain discretised with chosen unit-cell}(b) is made up of four particles with two degrees of freedom per particle. The vector of displacements for one unit-cell is $\mathbf{u} = [u_{A}, v_{A}, u_{B}, v_{B}, u_{C}, v_{C}, u_{D}, v_{D}]^{T}$. With the displacement vector $\mathbf{u}$ and the stiffness matrix $\mathbf{K}_{uc}^{mod}$ given in Appendix \ref{appendix: stiffness matrix of the unit-cell with improved bond model}, the strain energy of the unit-cell in Equation (\ref{eqn: strain energy in unit-cell with improved bond model}) can alternatively be written as
\begin{align} \label{eqn: strain energy of unit-cell quadratic form of displacements}
\Pi_{uc}^{mod} = \frac{1}{2} \mathbf{u}^{T} \mathbf{K}_{uc}^{mod} \mathbf{u}.	
\end{align}

\begin{figure}[H]
\centering
\subfigure[\textit{Born model} - plane stress]{\includegraphics[width=0.49\textwidth]{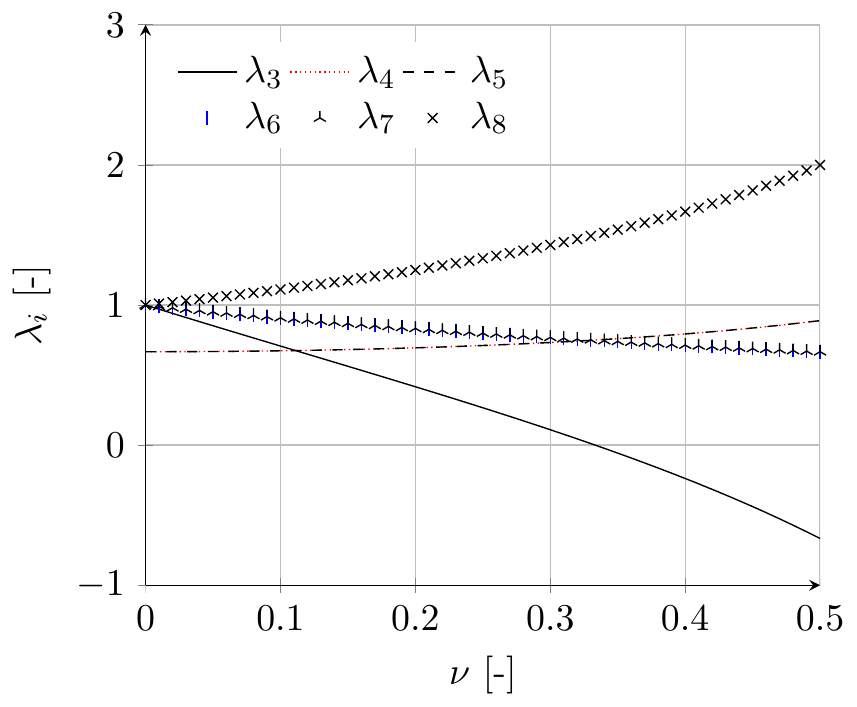}}
\hfill
\subfigure[Modified model - plane stress]{\includegraphics[width=0.49\textwidth]{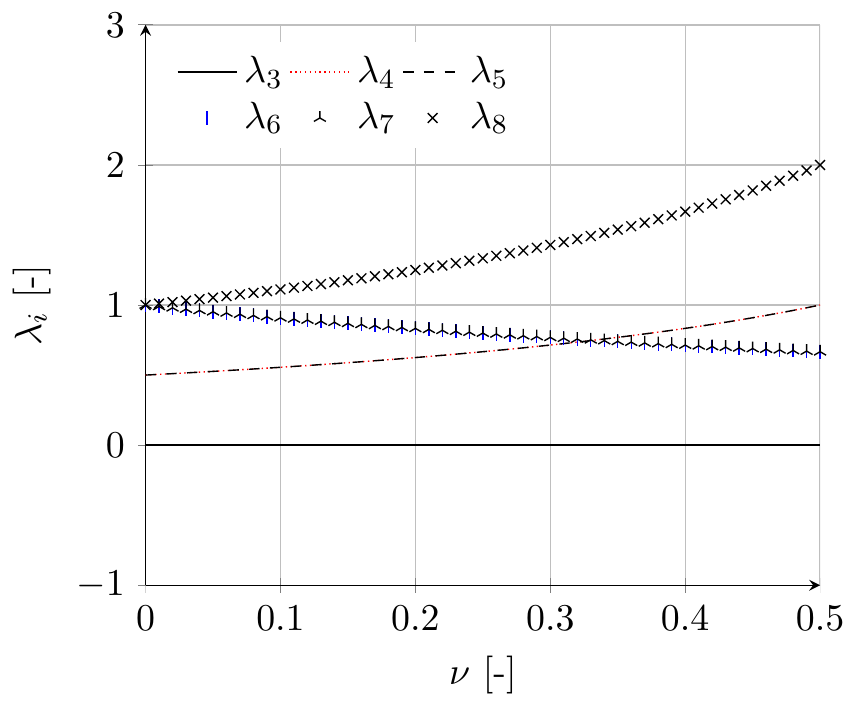}}
\vfill
\subfigure[\textit{Born model} - plane strain]{\includegraphics[width=0.49\textwidth]{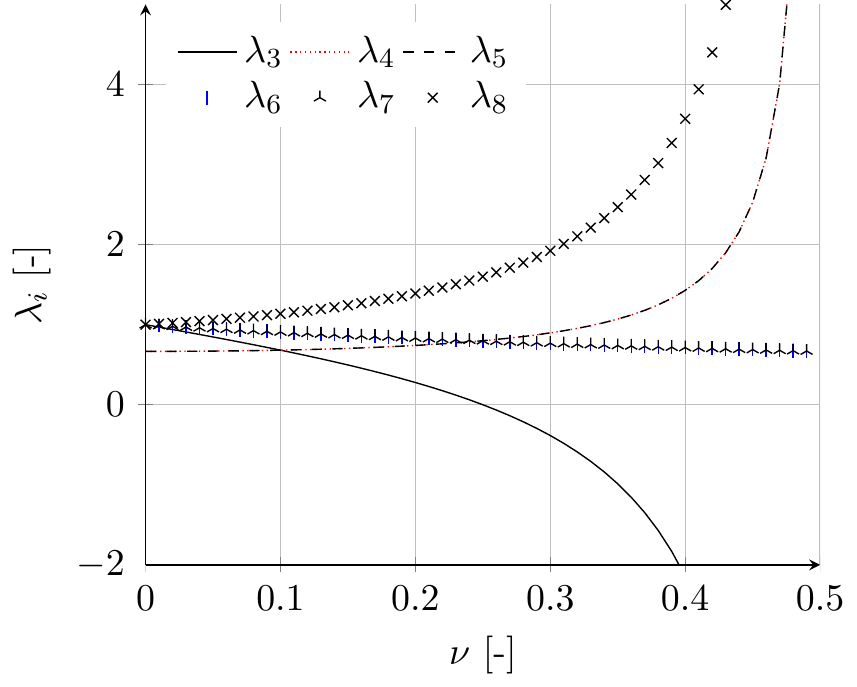}}
\hfill
\subfigure[Modified model - plane strain]{\includegraphics[width=0.49\textwidth]{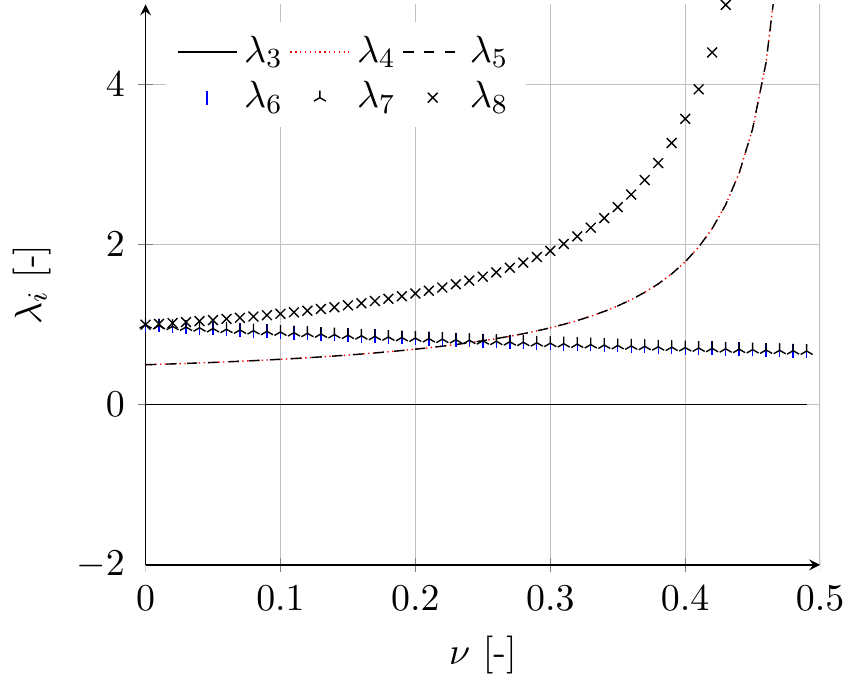}}
\caption{Variation of normalised eigenvalues of unconstrained unit-cell stiffness matrix as a function of $\nu$}
\label{fig: variation of eigenvalues of unit-cell}
\end{figure}

Equation (\ref{eqn: strain energy of unit-cell quadratic form of displacements}) shows that the strain energy of the unit-cell is a quadratic function of displacements. In order to be thermodynamically stable, it must be positive definite. The necessary and sufficient condition for the positive definiteness of Equation (\ref{eqn: strain energy of unit-cell quadratic form of displacements}) is that all eigenvalues of the eigenvalue problem $\mathbf{K}_{uc}^{mod} \mathbf{u} = \boldsymbol{\lambda} \mathbf{u}$ are real and positive \cite{Fung:1965aa}. The relating stiffness matrices are given in Appendix \ref{appendix: stiffness matrix of the unit-cell with Born bond model} and \ref{appendix: stiffness matrix of the unit-cell with improved bond model} respectively. The resulting eigenvalues and corresponding eigenforms are summarised in Table \ref{table: eigenvalues of unit-cell}. After substituting the stiffness parameters of the unit cell from Table \ref{table: calibrated stiffness parameters of unmodified and improved square unit-cell}, the non-zero eigenvalues are plotted as a function of Poisson's ratio $\nu$ for both plane stress and plane strain in Figure \ref{fig: variation of eigenvalues of unit-cell}. The stiffness parameters are normalised with respect to Young's modulus $E$ and the thickness $t$ to obtain a qualitative behaviour. 

\begin{table}[H]
\caption{Eigenvalues and eigenforms of unit-cell with the \textit{Born model} and modified model}
 \label{table: eigenvalues of unit-cell}
 \centering
\begin{tabular}[]{c c c c}\hline
\noalign{\smallskip}$\lambda_{i}$ & Eigenform & \textit{Born} model & Modified bond model\\ \noalign{\smallskip}\hline
\noalign{\smallskip}$\lambda_{1}$ & \raisebox{-.4\height}{\includegraphics[width=0.08\textwidth]{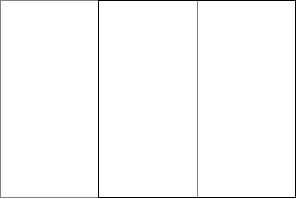}} & 0 & 0\\ \noalign{\smallskip}\hline
\noalign{\smallskip}$\lambda_{2}$ & \raisebox{-.4\height}{\includegraphics[height=0.08\textwidth]{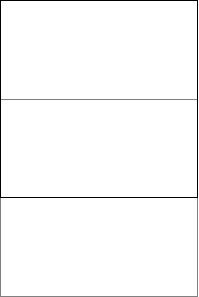}} & 0 & 0\\ \noalign{\smallskip}\hline
\noalign{\smallskip}$\lambda_{3}$ & \raisebox{-.4\height}{\includegraphics[width=0.08\textwidth]{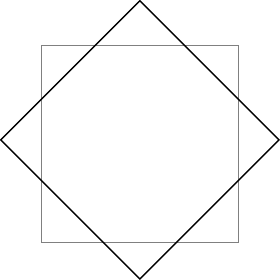}} & $3 k_{s_{1}}^{b}$ & 0\\ \noalign{\smallskip}\hline
\noalign{\smallskip}$\lambda_{4}$ & \raisebox{-.4\height}{\includegraphics[width=0.08\textwidth]{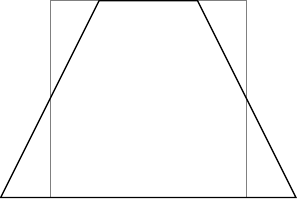}} & $k_{n_{1}}^{b} + k_{s_{1}}^{b}$ & $k_{n_{1}}^{m} + k_{s_{1}}^{m}$\\ \noalign{\smallskip}\hline
\noalign{\smallskip}$\lambda_{5}$ & \raisebox{-.4\height}{\includegraphics[height=0.08\textwidth]{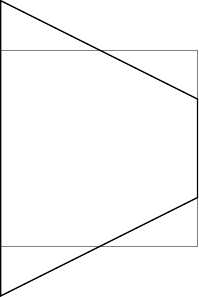}} & $k_{n_{1}}^{b} + k_{s_{1}}^{b}$ & $k_{n_{1}}^{m} + k_{s_{1}}^{m}$\\ \noalign{\smallskip}\hline
\noalign{\smallskip}$\lambda_{6}$ & \raisebox{-.4\height}{\includegraphics[width=0.08\textwidth]{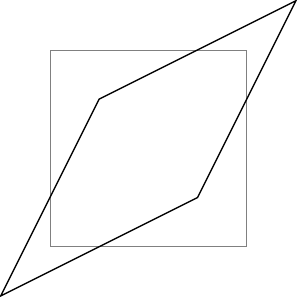}} & $2 k_{n_{2}}^{b} + k_{s_{1}}^{b}$ & $2 k_{n_{2}}^{m} + 2 k_{s_{1}}^{m}$\\ \noalign{\smallskip}\hline
\noalign{\smallskip}$\lambda_{7}$ & \raisebox{-.4\height}{\includegraphics[width=0.08\textwidth]{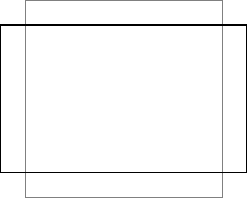}} & $k_{n_{1}}^{b} + 2 k_{s_{1}}^{b}$ & $k_{n_{1}}^{m} + 4 k_{s_{1}}^{m}$\\ \noalign{\smallskip}\hline
\noalign{\smallskip}$\lambda_{8}$ & \raisebox{-.4\height}{\includegraphics[width=0.08\textwidth]{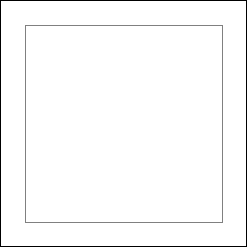}} & $k_{n_{1}}^{b} + 2 k_{n_{2}}^{b}$ & $k_{n_{1}}^{m} + 2 k_{n_{2}}^{m}$\\ \noalign{\smallskip}\hline
\end{tabular}
\end{table}

From Figure \ref{fig: variation of eigenvalues of unit-cell} the following observations are made for both plane stress and plane strain:
\begin{enumerate}
	\item Regarding the \textit{Born model}, the eigenvalue $\lambda_{3}$ of the rigid-body rotation eigenform is non-zero everywhere expect at $\nu = 1/3$ and $\nu = 1/4$ for plane stress and plane strain respectively. This implies that the \textit{Born model} deforms under rigid-body rotation and therefore it is proven that this model cannot distinguish between rigid-body rotation and shear deformations. Also, the eigenvalue becomes negative for $\nu > 1/3$ and $\nu > 1/4$ for plane stress and plane strain respectively. This implies that the strain energy function derived from the \textit{Born model} is not positive definite after this lower limit of the Poisson's ratio. Therefore we expect the response to be unstable for values of Poisson's ratio above this limit. However, regarding the modified bond model, the eigenvalue of the rigid-body rotation eigenform remains zero for all values of $\nu$ and all eigenvalues remain positive. Hence, the strain energy derived from the modified bond model is positive definite and therefore is stable for all values of $\nu$ although the shear stiffness $k_{s_{1}}^{m}$ of the unit-cell takes on a negative value after a critical value of $\nu$ as shown in Figure \ref{fig: plot of calibrated stiffness parameters and elasticity tensor components}(b). This implies that a negative stiffness does not necessarily lead to unstable results, rather it is due to negative eigenvalues. Similarly, in the work of Esin \cite{Esin:2016aa} it was concluded that materials and structures with negative stiffness elements can exist when the negative element energy is compensated by the energy of the rest of the system or an encompassing system that provides stabilisation.
	\item The eigenvalues ($\lambda_{4}$, $\lambda_{5}$) related to bending eigenforms result as expected for an isotropic material.
	\item Similarly, the eigenvalues of shear ($\lambda_{6}$, $\lambda_{7}$) eigenforms result as expected.
	\item As expected, the volumetric form $\lambda_{8}$ has the highest eigenvalue.
\end{enumerate}

\paragraph{Remark:} At $\nu = 1/3$ and $\nu = 1/4$ for plane stress and plane strain, the shear stiffness becomes zero and the model reduces to one where it has only normal springs. At this value, the \textit{Born model} is rotationally invariant and hence we have $\lambda_{3} = 0$ at this particular value of $\nu$.

\section{Validation examples}
\label{sec: validation examples}
In order to validate the capability of the modified model and compare it to the \textit{Born model}, four different examples are chosen for which closed-form continuum solutions exist. With the help of these examples and based on the positive definiteness of the strain energy function, the stability of the bond model will also be investigated. The discrete element method results are obtained by solving the linear system of equations $\mathbf{K} \, \mathbf{u} = \mathbf{f}$ for the unknown particle displacements $\mathbf{u}$. The system stiffness matrix $\mathbf{K}$ is obtained by assembling the individual unit-cell stiffness matrices derived in the Appendix \ref{appendix: stiffness matrix of the unit-cell with Born bond model} and \ref{appendix: stiffness matrix of the unit-cell with improved bond model}. External tractions are reformulated as forces and are added to the corresponding position in the global force vector $\mathbf{f}$. 

\subsection{Uniaxial}
\label{subsec: uniaxial}
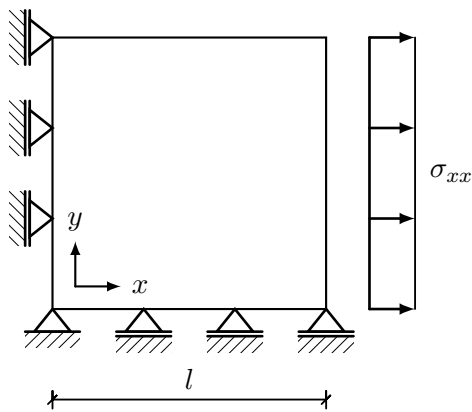
\begin{figure}[H]
\centering
\input{tikz/figure7}
\caption{Thin plate under uniaxial tension}
\label{fig: uniaxial tension}
\end{figure}

A thin square plate of length $l = \SI{0.2}{[m]}$ and thickness $t = \SI{0.01}{[m]}$ as shown in Figure \ref{fig: uniaxial tension} is modelled with the chosen square unit-cell under plane stress conditions with $E = \SI{2e11}{[\newton\per\square\metre]}$. The plate is constrained in $y-$direction along the bottom and in the $x-$direction along the left. A stress $\sigma_{xx}$ is applied on the right edge to model a uniaxial state of deformation. 
The analytical solution for the displacement fields is $u = \frac{\sigma_{xx}}{E} x$ and $v = -\frac{\nu \, \sigma_{xx}}{E}y$. The resulting displacements along $u(x=l, y)$ and $v(x, y=l)$ for three different values of Poisson's ratio $\nu$ are summarised in the Table \ref{table: comparison of uniaxial tension test results} for both the bond models. 

\begin{figure}[H]
\centering
\subfigure[\textit{Born model}]{\includegraphics[width=0.49\textwidth]{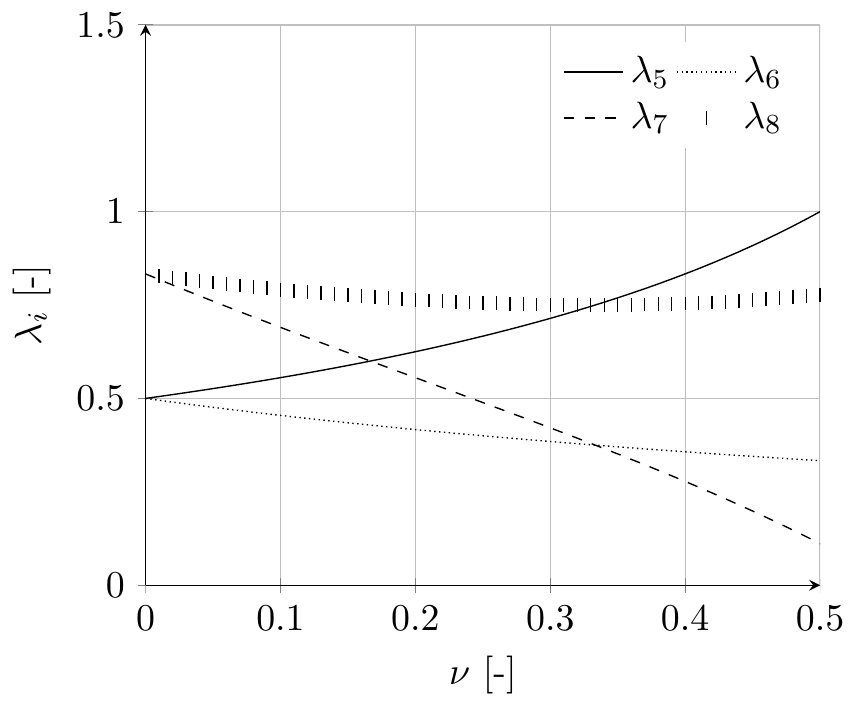}}
\subfigure[Modified model]{\includegraphics[width=0.49\textwidth]{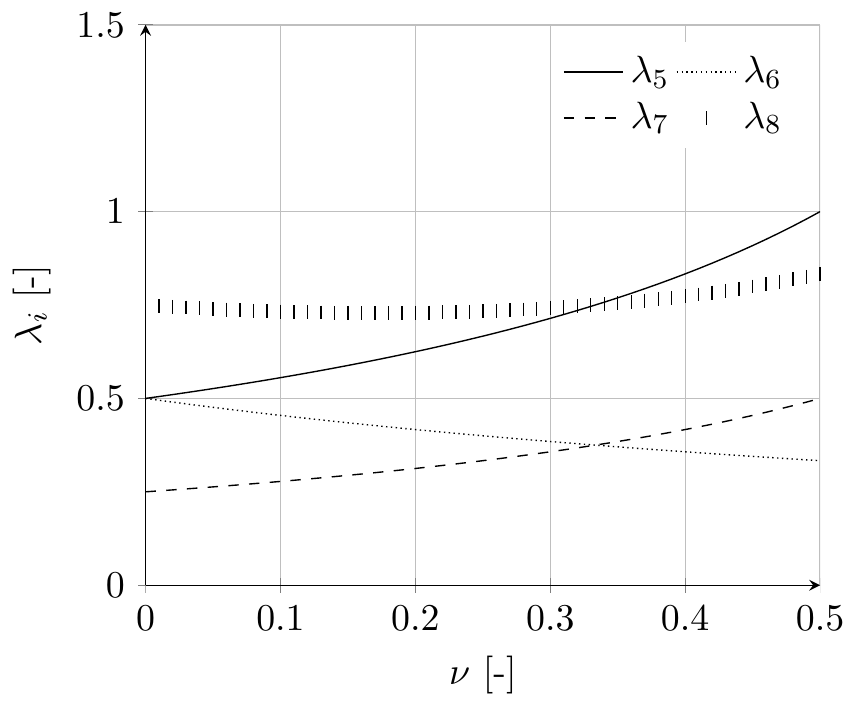}}
\caption{Variation of eigenvalues of constrained unit-cell stiffness matrix as a function of $\nu$ for uniaxial test}
\label{fig: variation of uniaxial constrained eigenvalues}
\end{figure}

We observe that both unit-cells employing the \textit{Born model} and the modified bond model produce the exact results for all values of  the Poisson's ratio independent of the discretisation, because the analytical solution is a linear function of the position and both bond models use linear springs. Although both models employ negative shear stiffness for $\nu > 1/3$, they still produce the expected result. In Section \ref{subsec: positive definiteness of strain energy}, based on the eigenvalues of the stiffness matrix $\mathbf{K}_{uc}$ of an unconstrained unit-cell employing the \textit{Born model} it was observed that the strain energy was not positive definite (due to the rigid-body rotation eigenform) for $\nu > 1/3$ in the case of plane stress. Therefore we may expect unstable results. However due to the constraints used in the example, all rigid-body eigenforms vanish and only positive eigenvalues exists for all values of $\nu$. This is shown in Figure \ref{fig: variation of uniaxial constrained eigenvalues}, where the normalised eigenvalues of the constrained unit-cell employing the \textit{Born model} and the modified model are plotted as a function of $\nu$.

\begin{table}
\centering
\caption{Comparison of uniaxial tension test results (Displacements in $\si{[\times 10^{-3} \meter]}$)}
\label{table: comparison of uniaxial tension test results}
\begin{tabular}[]{c c c c c c c}
\hline\noalign{\smallskip}
$\nu$ & \multicolumn{2}{c}{Analytical solution} & \multicolumn{2}{c}{\textit{Born model}} & \multicolumn{2}{c}{Modified model}\\ \noalign{\smallskip} \hline
\noalign{\smallskip}$\phantom{0}$ & $u$ & $v$ & $u$ & $v$ & $u$ & $v$\\ \noalign{\smallskip}\cline{2-7}
\noalign{\smallskip}$0$ & $0.1$ & $0$ & $0.1$ & $0$ & $0.1$ & $0$\\ \noalign{\smallskip}\hline
\noalign{\smallskip}$0.3$ & $0.1$ & $-0.03$ & $0.1$ & $-0.03$ & $0.1$ & $-0.03$\\ \noalign{\smallskip}\hline
\noalign{\smallskip}$0.49$ & $0.1$ & $-0.049$ & $0.1$ & $-0.049$ & $0.1$ & $-0.049$\\ \noalign{\smallskip}\hline
\end{tabular}
\end{table}

\subsection{Pure shear}
\label{subsec: pure shear}
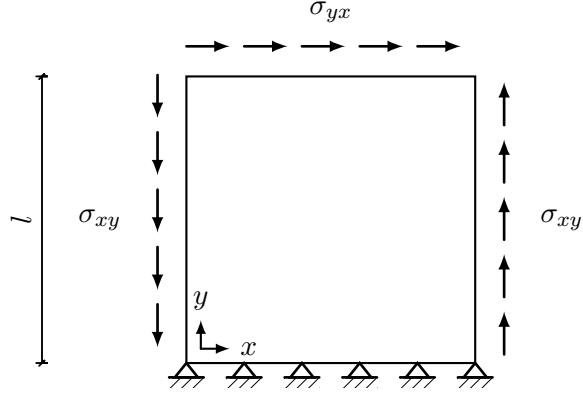
\begin{figure}
\centering
\input{tikz/figure9}		
\caption{Thin plate under pure shear}
\label{fig: pure shear}
\end{figure}

As a second test, a thin square plate with edge length $l = \SI{0.2}{[\meter]}$ and thickness $t = \SI{0.01}{[\meter]}$ as shown in Figure \ref{fig: pure shear} is modelled under plane stress condition with $E = \SI{2e11}{[\newton\per\square\metre]}$ applying pure shear stress $\sigma_{xy} = \sigma_{yx} = \tau_{0} = \SI{1e8}{[\newton\per\square\metre]}$.

\begin{figure}[H]
\centering
\scriptsize
\subfigure[\textit{Born model}: $\nu = 0$]{\def\svgwidth{0.95\textwidth} 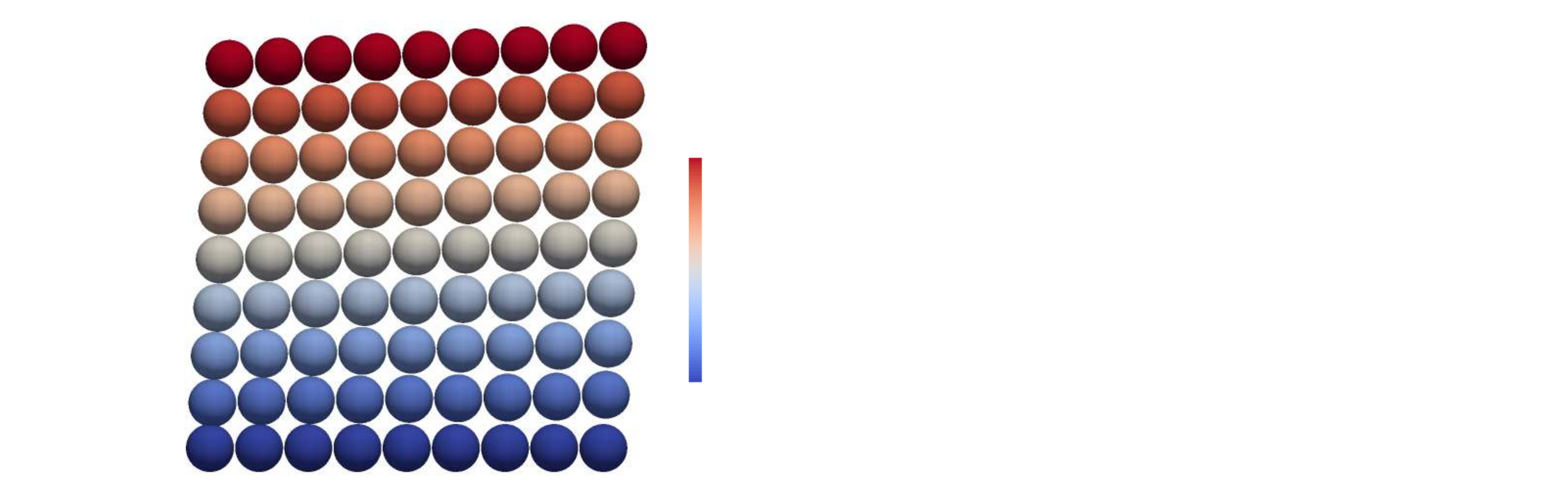}
\vfill
\subfigure[\textit{Born model}: $\nu = 0.3$]{\def\svgwidth{0.95\textwidth} 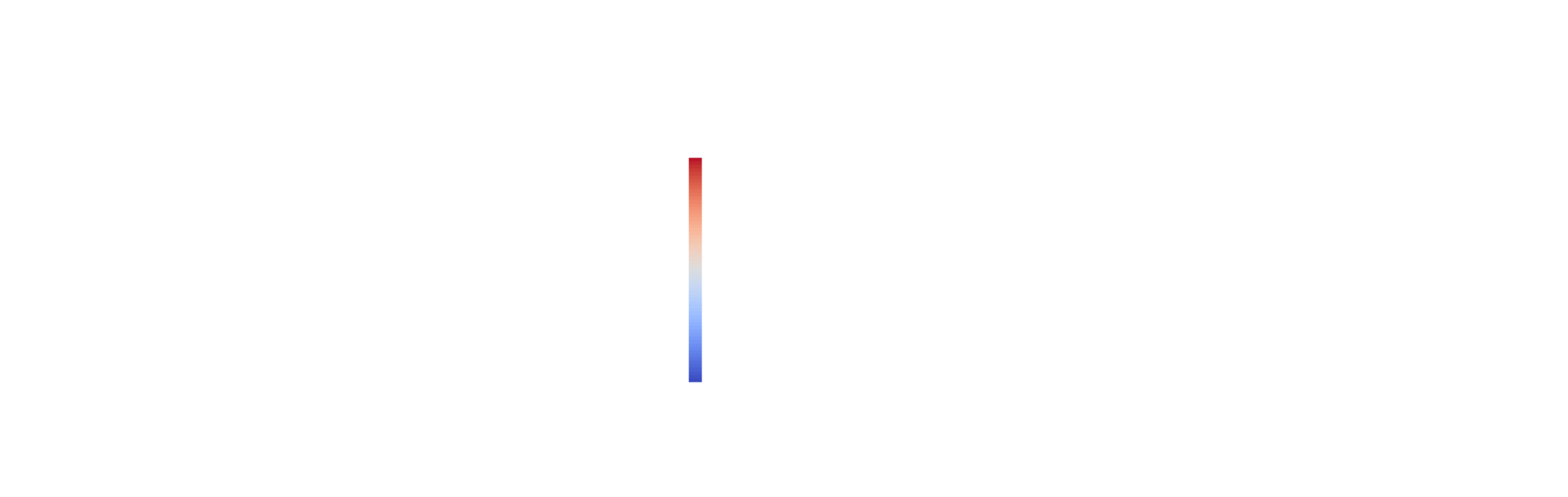}
\vfill
\subfigure[\textit{Born model}: $\nu = 0.49$]{\def\svgwidth{0.95\textwidth} 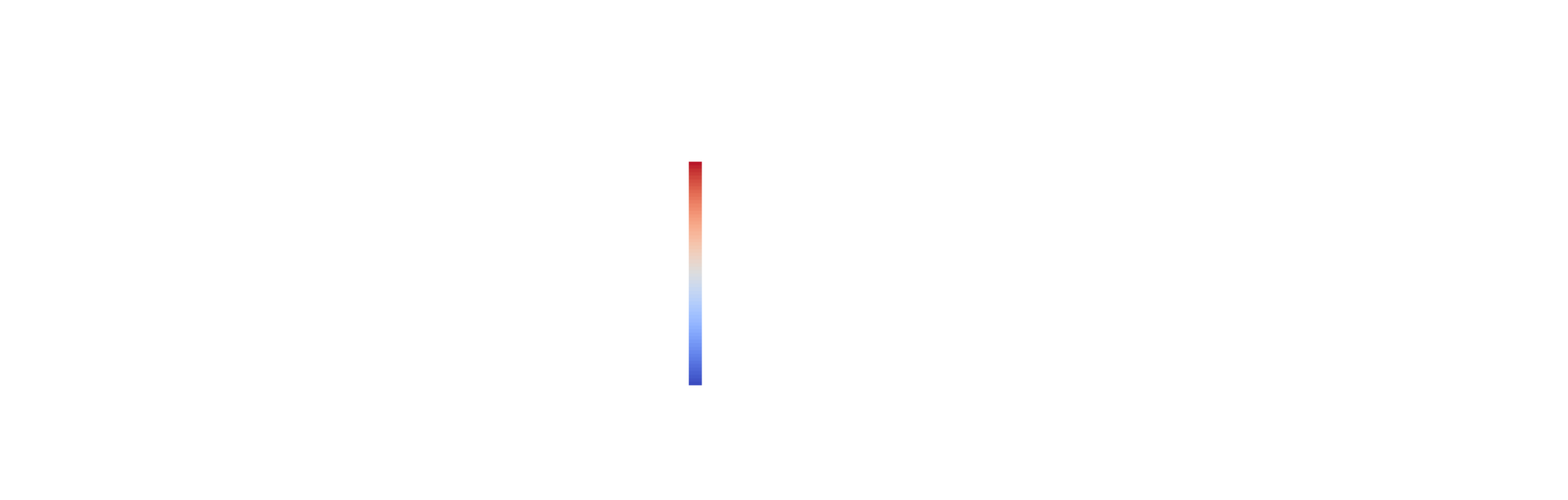}
\vfill
\subfigure[Modified model: $\nu = 0.49$]{\def\svgwidth{0.9\textwidth} 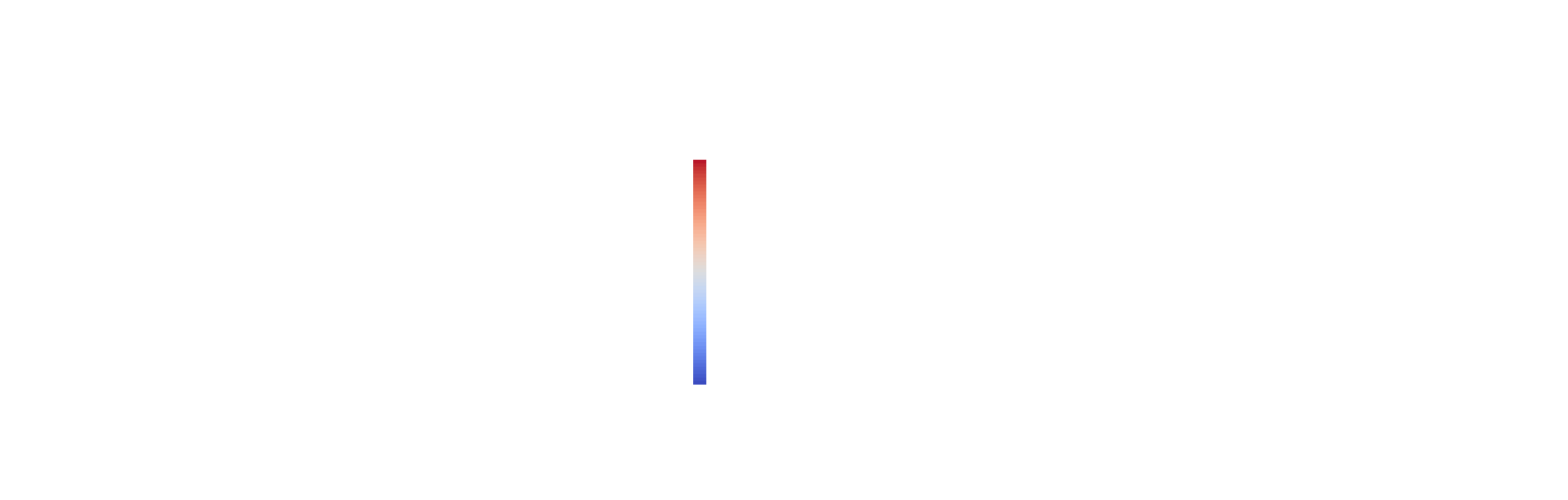}
\caption{Displacements $u$ (left) and $v$ (right) employing the \textit{Born model} and the modified model}
\label{fig: pure shear Born and modified model}
\end{figure}

The plate is constrained in both directions along the bottom edge to prevent rigid-body motion similar to that used by Ockelmann \cite{Ockelmann:2017aa}. The analytical solution for the displacement fields is obtained as $u = \frac{\tau_{0}}{G} y$ and $v = 0$, where $G = E/2(1 + \nu)$ is the shear modulus. Here again, the test is conducted for three different values of $\nu$ and also for four different discretisations similar to the uniaxial test. The $u$ and $v$ displacements of the domain discretised with $8 \times 8$ unit-cells employing the \textit{Born model} under pure shear are summarised in Figure \ref{fig: pure shear Born and modified model}(a), (b) and (c). In Figure \ref{fig: pure shear Born and modified model}(d), the results of $8 \times 8$ unit-cells employing the modified model under pure shear for $\nu = 0.49$ are given. In Figure \ref{fig: pure shear Born and modified model} the bonds connecting the particles are not visualised for clarity. 

\begin{figure}[H]
\begin{minipage}[t]{0.5\textwidth}
\includegraphics[width=\textwidth]{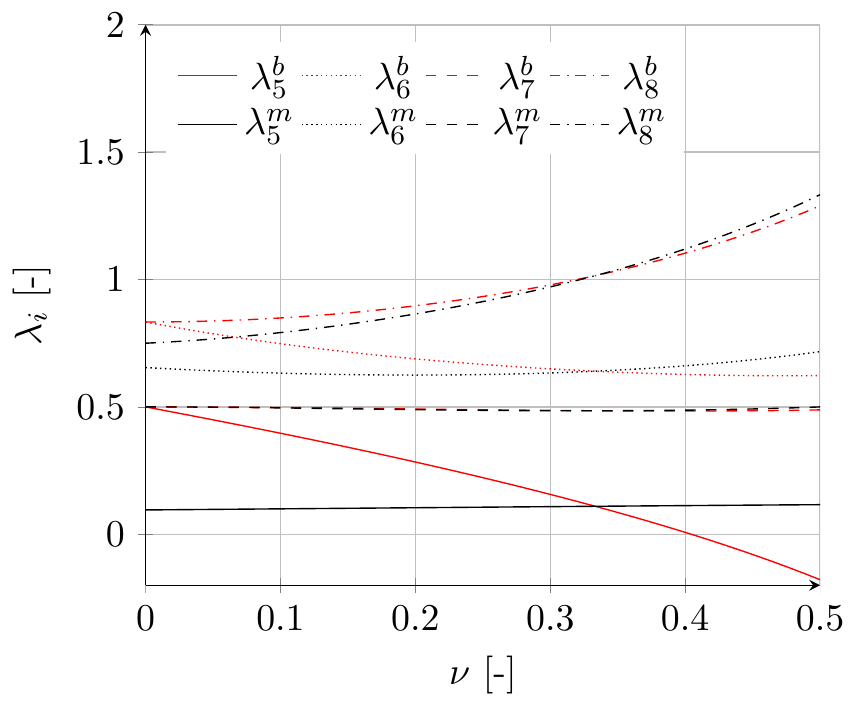}
\caption{Variation of eigenvalues of constrained unit-cell stiffness matrix as a function of $\nu$ for pure shear test}
\label{fig: variation of shear constrained eigenvalues}
\end{minipage}
\hspace{5mm}
\begin{minipage}[t]{0.45\textwidth}
\includegraphics[width=0.95\textwidth]{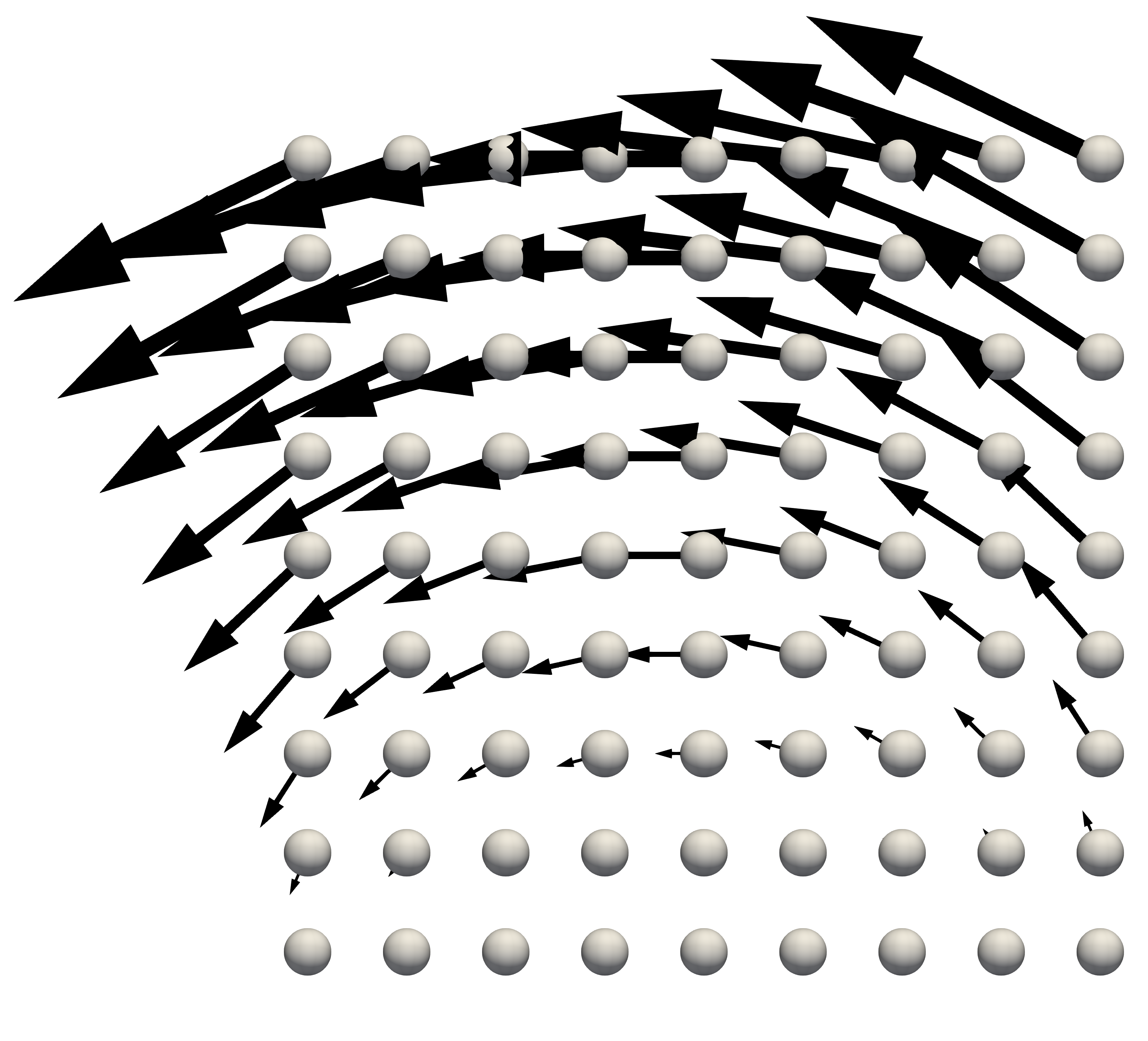}
\caption{Eigenform corresponding to $\lambda_{5}$}
\label{fig: eigenform corresponding to lambda5}
\end{minipage}
\end{figure}

We observe that the results for the unit-cell employing the modified model agrees exactly with the analytical solution, because similar to the case of uniaxial tension, the analytical displacement field is a linear function of the position of particles. With the modified bond model, the exact solution was also observed for all the discretisations. However, for the unit-cell employing the \textit{Born model} the response of the plate is stiffer in comparison with the analytical solution in the range $0 < \nu \leq 1/3$ and unstable for $\nu > 1/3$. In order to understand the reason for this behaviour, the plate is discretised with one unit-cell employing the \textit{Born model} and the modified model respectively. The normalised eigenvalues of the constrained stiffness matrix are plotted as function of $\nu$ as shown in Figure \ref{fig: variation of shear constrained eigenvalues}. For the stiffness matrix of an unit-cell employing the \textit{Born model}, we observe from Figure \ref{fig: variation of shear constrained eigenvalues} that the eigenvalue $\lambda_{5}^{b}$ decreases with increasing values of $\nu$ and also becomes negative for $\nu > 0.4$. The eigenform corresponding to this eigenvalue is plotted as given as a quiver plot in Figure \ref{fig: eigenform corresponding to lambda5}. However, as the plate is further discretised, the eigenvalues of the unit-cells that are not located at the bottom edge of the plate take on the form as previously shown in Figure \ref{fig: variation of eigenvalues of unit-cell}(a) for the case of an unconstrained unit-cell stiffness matrix employing the Born model. This unconstrained stiffness matrix also includes the contribution of rigid-body rotation and is unstable for values of $\nu > 1/3$ for 

the case of plane stress. Therefore, the upper limit on the Poisson's ratio for the plate constrained as shown in Figure \ref{fig: pure shear} is $\nu = 1/3$ employing the \textit{Born model}. However, for the unit-cell employing the modified model, due to the coupling of shear strain energy of neighbouring bonds, the eigenvalues remains positive for all values of $\nu$ as shown in Figure \ref{fig: variation of shear constrained eigenvalues} for the constrained stiffness-matrix and in Figure \ref{fig: variation of eigenvalues of unit-cell}(b) for the unconstrained stiffness matrix respectively.

\subsection{Pure bending}
\label{subsec: pure bending}
\begin{figure}[H]
	\centering
	\input{tikz/figure13}
	\caption{Thin plate under pure bending}
	\label{fig: pure bending}
\end{figure}
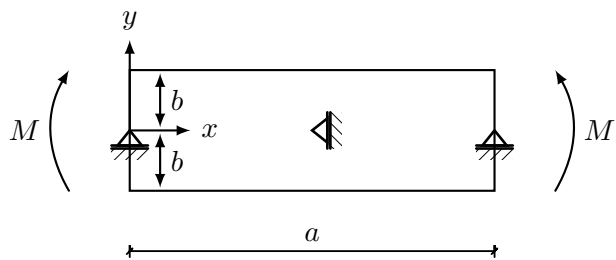

Next, a thin rectangular plate of length $l = a = \SI{0.5}{[m]}$, height $h = 2b = \SI{0.125}{[m]}$ and thickness $t = \SI{0.01}{[m]}$ under pure bending is considered. The plate is constrained at two points in the $y-$direction and at one location in the middle along $x-$direction as shown in Figure \ref{fig: pure bending}. The applied bending moment $M = \SI{2604.17}{[\newton\metre]}$ at the sides is modelled as a linearly varying load $\sigma_{0}$ with opposite magnitude at both corners. Following Timoshenko \cite{Timoshenko:1969aa}, the analytical displacement fields are $u = \frac{M y}{EI}(-x + \frac{l_{x}}{2})$ and $v = \frac{M}{2EI}(\nu y^{2} + x^{2} - x l_{x})$, where $I = th^{3}/12$.

\vfill

\begin{figure}
\centering
\subfigure[\textit{Born model}]{\includegraphics[width=0.49\textwidth]{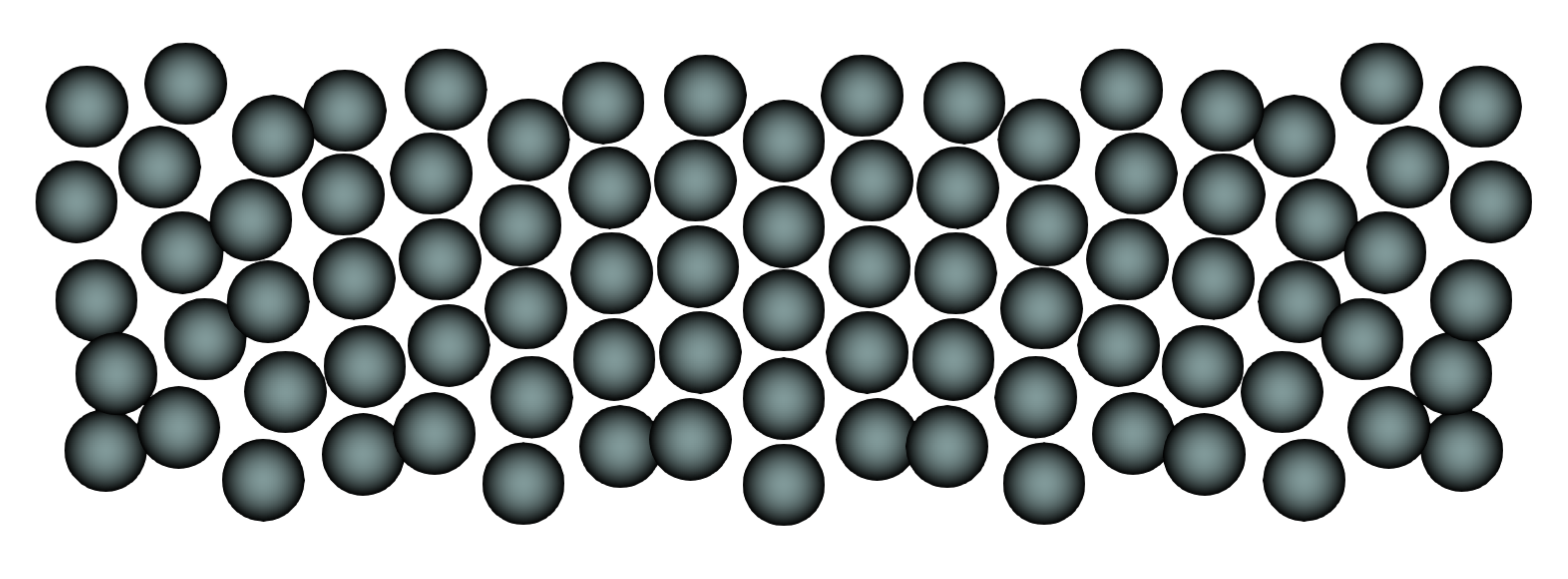}}
\subfigure[Modified model]{\includegraphics[width=0.49\textwidth]{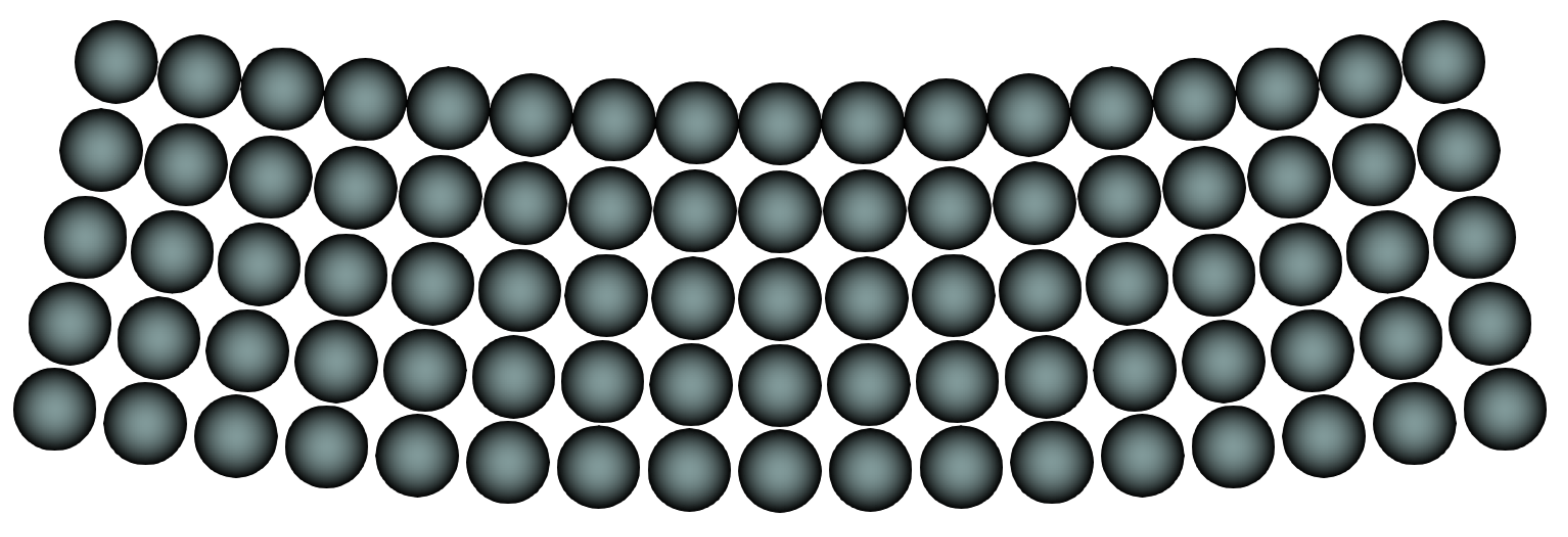}}
\caption{Deformed configuration for $\nu = 0.49$}
\label{fig: pure bending deformed configuration}
\end{figure}

The test was performed with four different discretisations: $8 \times 2$, $16 \times 4$, $32 \times 8$ and $64 \times 16$. Also, the effect of $\nu$ on the accuracy has been checked by performing the test for three different values of $\nu$. The deformed configuration employing the \textit{Born model} and the modified model for $\nu = 0.49$ is given in Figure \ref{fig: pure bending deformed configuration}.

\begin{figure}
\centering
\includegraphics[width=0.8\textwidth]{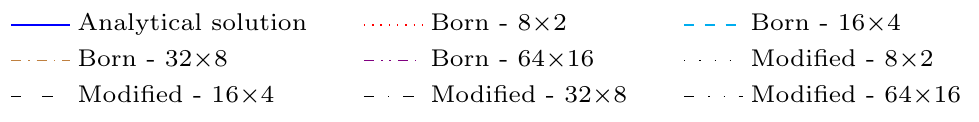}
\vfill
\subfigure[$\nu = 0$]{\includegraphics[width=0.4\textwidth]{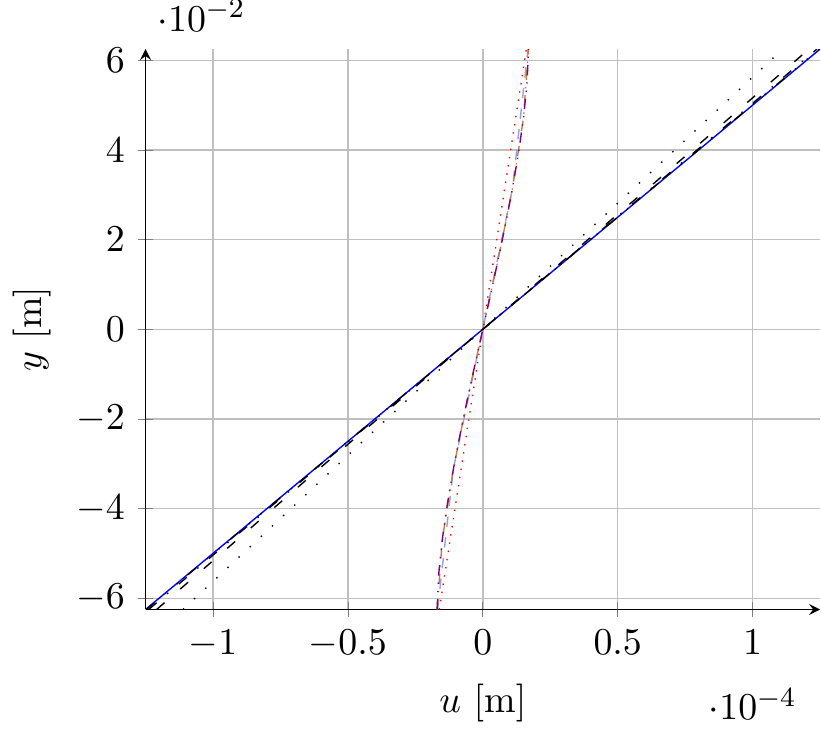}}
\hfil
\subfigure[$\nu = 0$]{\includegraphics[width=0.4\textwidth]{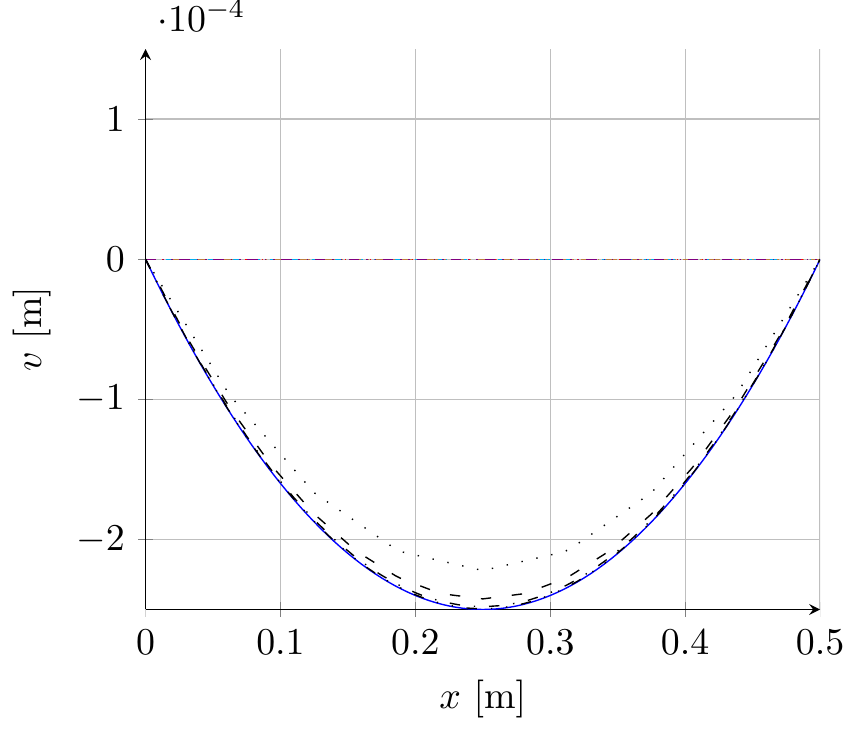}}
\vfill
\subfigure[$\nu = 0.3$]{\includegraphics[width=0.4\textwidth]{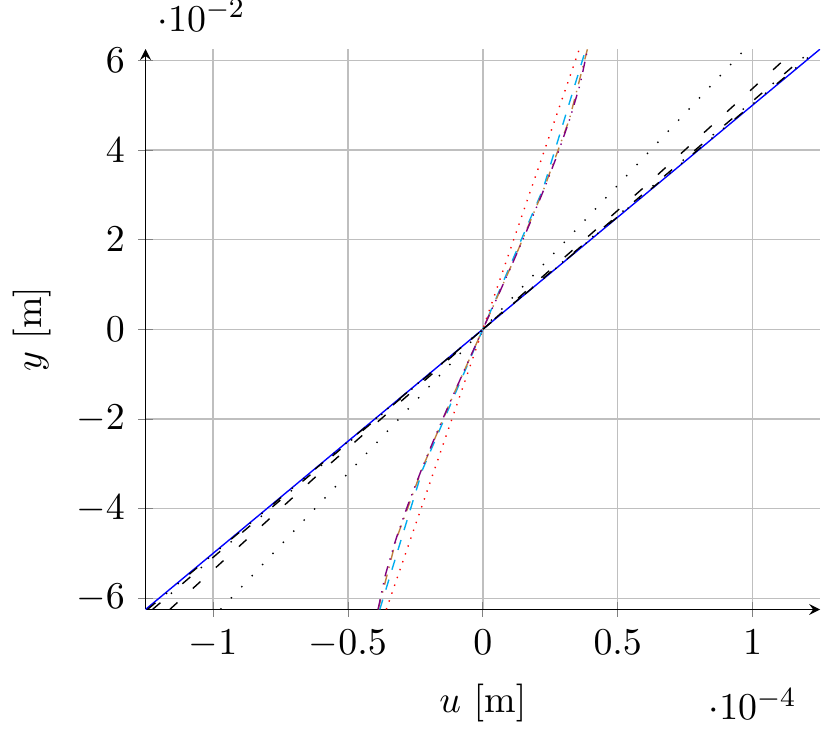}}
\hfil
\subfigure[$\nu = 0.3$]{\includegraphics[width=0.4\textwidth]{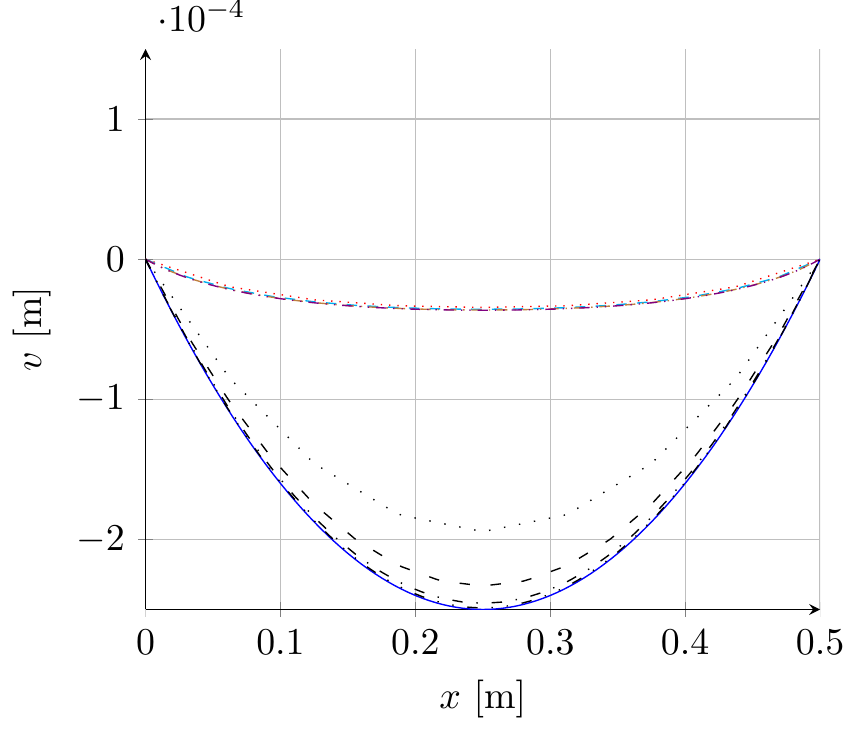}}
\vfill
\subfigure[$\nu = 0.49$]{\includegraphics[width=0.4\textwidth]{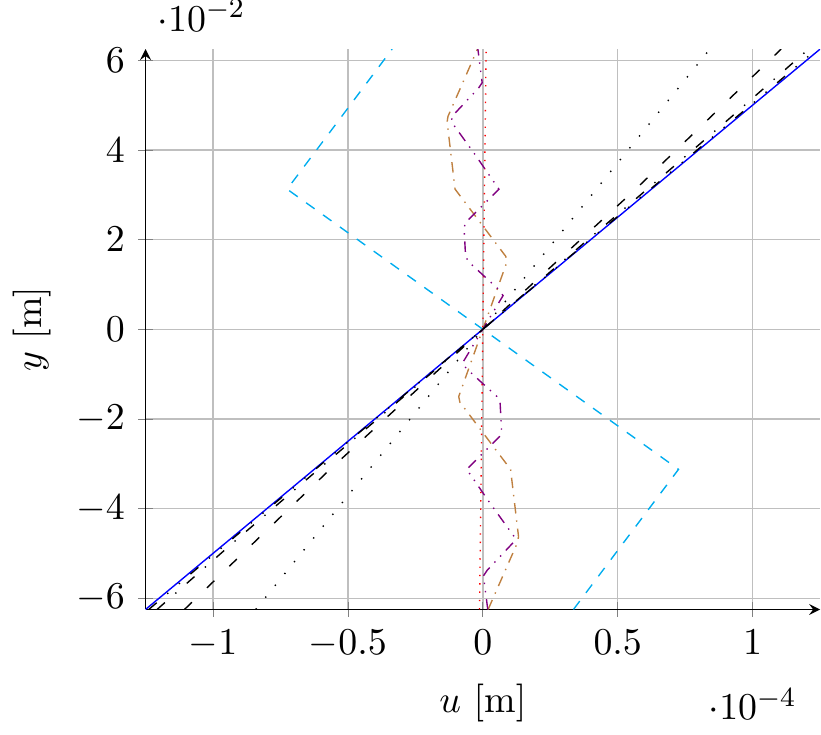}}
\hfil
\subfigure[$\nu = 0.49$]{\includegraphics[width=0.4\textwidth]{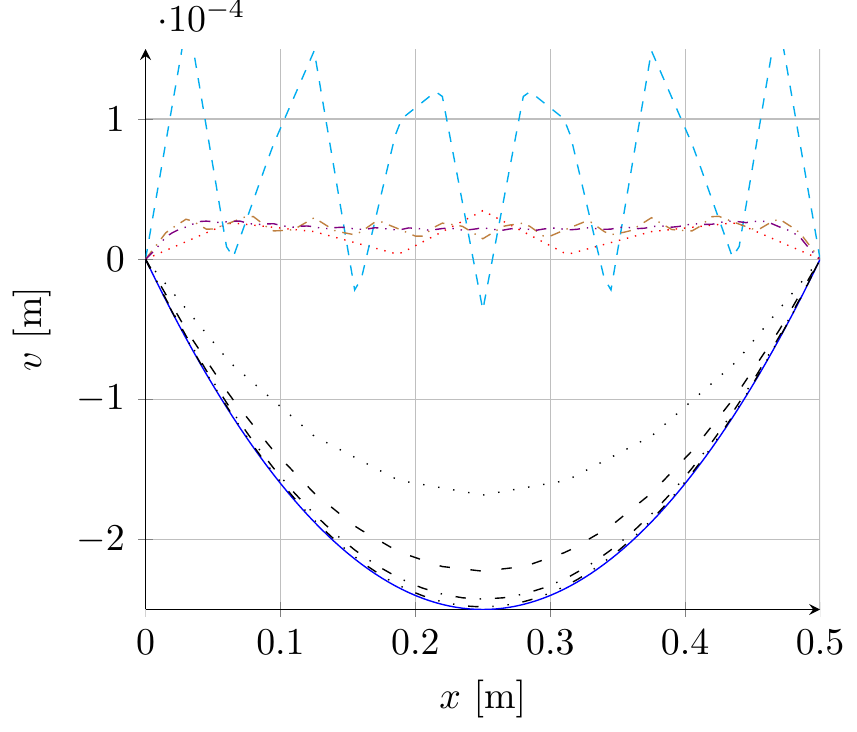}}
\caption{Displacements $u$ at the left edge of the plate (left)  and $v$ of the axis of the plate (right) employing the \textit{Born model} and the modified model}
\label{fig: comparison of displacements pure bending}
\end{figure}

The displacement $u$ at the left edge of the plate and the displacement $v$ of the axis of the plate employing the \textit{Born model} and the modified model are compared with the analytical solution for different discretisations and for different Poisson's ratio and are summarised in Figure \ref{fig: comparison of displacements pure bending}. From the results we observe that regarding the \textit{Born model} for $\nu = 0$ and $\nu = 0.3$, the displacements $u$ and $v$ are much smaller in comparison with that of the analytical solution. In the case of pure bending, elements which are located faraway from the boundary undergo rigid body rotations as well. Since the \textit{Born model} cannot distinguish between rigid-body rotations and shear deformations, strain energy is also stored for the rigid body rotation. As the rigid-body rotation eigenvalue takes on a maximum value at $\nu = 0$ and approaches zero at $\nu = 1/3$ for the chosen unit-cell employing the \textit{Born model}, the response of the plate is much stiffer for $\nu = 0$ than for $\nu = 1/3$. And for $\nu > 1/3$ the results are unstable since the eigenvalue takes a negative value and hence the strain energy function is no longer positive definite. We also observe regarding the unit-cell employing the \textit{Born model}, that the displacements do not converge. This is due to the non-zero eigenvalue of the rigid-body rotation eigenform. For the unit-cell employing the modified bond model, the coupling of the shear strain energy of the neighbouring bonds allows the model to distinguish between rigid-body rotations and shear deformations. Therefore, satisfactory results are obtained employing this modified bond model and also convergence to the analytical solution upon refinement is observed.

\newpage
\subsection{Cantilever bending}
\label{subsec: cantilever bending}
As a final test, a thin rectangular plate under cantilever bending is considered. The right edge of the plate is completely constrained and a constantly distributed load $F$ is applied on the left edge as shown in Figure \ref{fig: cantilever bending}. The plate is modelled with the chosen unit-cell under plane stress conditions. 

\begin{figure}[H]
    \begin{minipage}[H]{0.5\textwidth}
    \begin{figure}[H]
    	\input{tikz/figure16}
    \end{figure}	
    \vfill
    \end{minipage}
    \begin{minipage}[H]{0.5\textwidth}
    \begin{align*}
    	a &= l = \SI{0.5}{[m]}\\
    	b &= h/2 = \SI{0.0625}{[m]}\\
    	t &= \SI{0.01}{[m]}\\
    	E &= \SI{2e11}{[\newton\per\square\metre]}\\
    	F &= \SI{1.25e7}{[\newton\per\metre]}
    \end{align*}
    \end{minipage}
    \caption{Thin plate under cantilever bending}
    \label{fig: cantilever bending}
\end{figure}

From \cite{Barber:2009aa}, the analytical solution for the displacement fields are obtained as
\begin{align} \label{eqn: analytical solution cantilever bending}
u &= \phantom{+}\frac{3Fx^{2}y}{4Eb^{3}} + \frac{3F(1 + \nu)y}{2Eb} - \frac{F(2 + \nu)y^{3}}{4Eb^{3}} - \frac{3Fa^{2}y}{4Eb^{3}} - \frac{3Fa^{2}y}{4Eb^{3}} \, \bigg(1 + \frac{(8 + 9\nu)b^{2}}{5a^{2}}\bigg)\\
v &=	-\frac{3F \nu xy^{2}}{4Eb^{3}} - \frac{Fx^{3}}{4Eb^{3}} - \frac{Fa^{3}}{2Eb^{3}} \, \bigg(1 + \frac{(12 + 11\nu)b^{2}}{5a^{2}}\bigg) + \frac{3Fa^{2}x}{4Eb^{3}} \, \bigg(1 + \frac{(8 + 9\nu)b^{2}}{5a^{2}}\bigg).
\end{align}
For the derivation of the analytical solution the following weak boundary conditions were applied along the right edge ($x = a$):
\begin{align} \label{eqn: boundary condition cantilever bending}
\int\limits_{-b}^{b} u \, dy = 0, \hspace{1cm} \int\limits_{-b}^{b} v \, dx = 0 \hspace{5mm} \text{and} \hspace{5mm} \int\limits_{-b}^{b} y \, v \, dy = 0.
\end{align}

\begin{figure}
\centering
\subfigure[\textit{Born model}]{\includegraphics[width=0.49\textwidth]{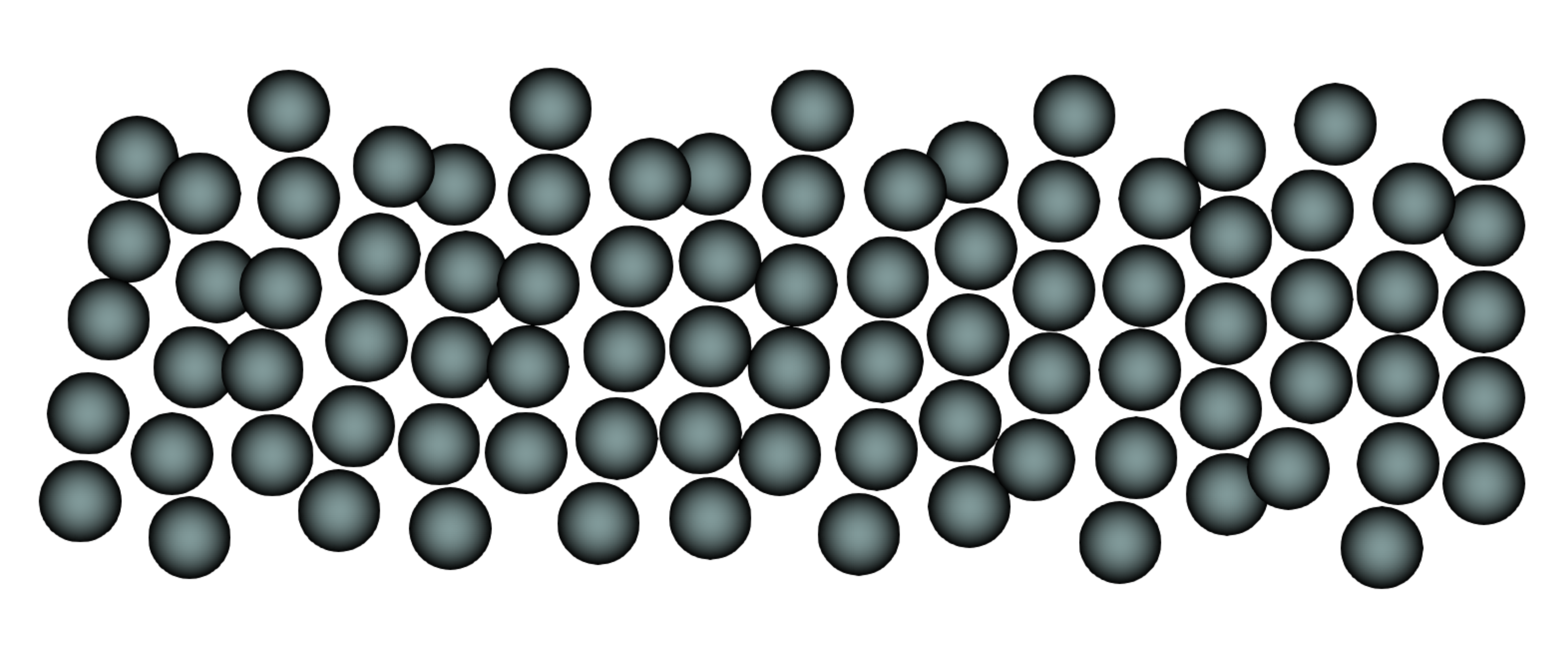}}
\subfigure[Modified model]{\includegraphics[width=0.49\textwidth]{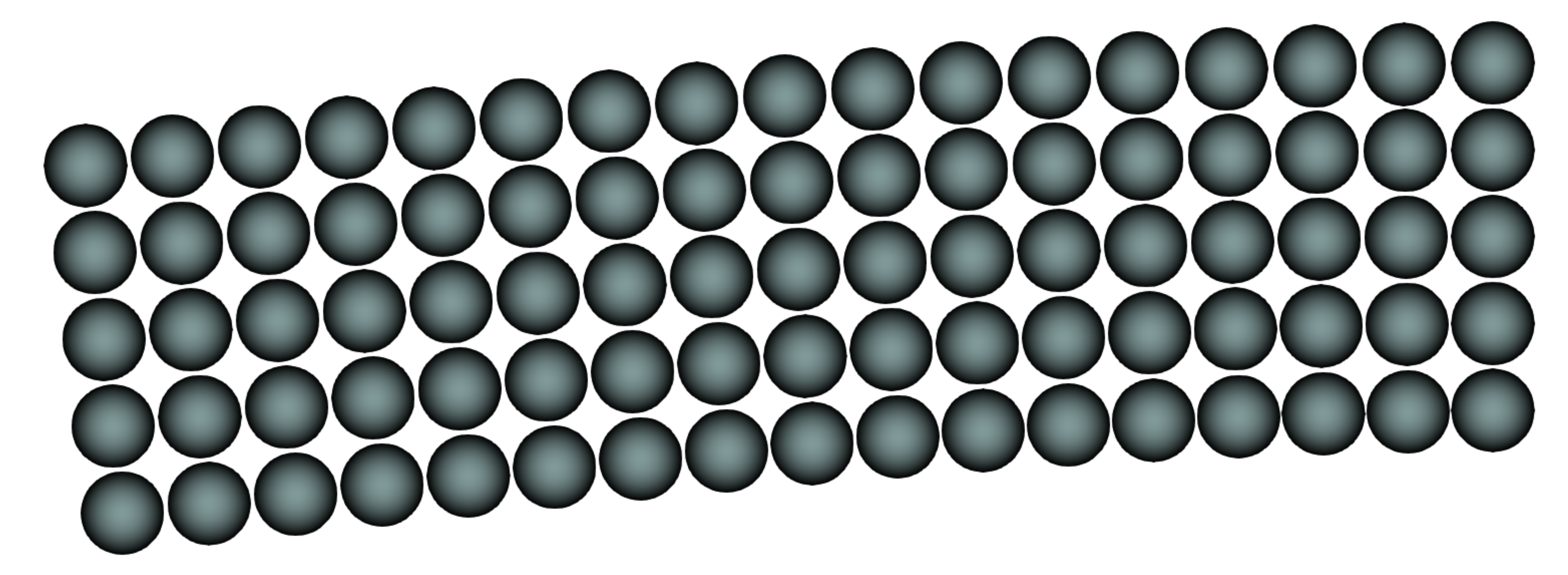}}
\caption{Deformed configuration for $\nu = 0.49$}
\label{fig: cantilever bending deformed configuration}
\end{figure}

Similar to the pure bending test, four different discretisations are used. The effect of $\nu$ on the results is studied as well. The deformed configuration employing the \textit{Born model} and the modified model are given in Figure \ref{fig: cantilever bending deformed configuration}. 

\begin{figure}
\centering
\includegraphics[width=0.8\textwidth]{figures/Legend.pdf}
\vfill
\subfigure[$\nu = 0$]{\includegraphics[width=0.4\textwidth]{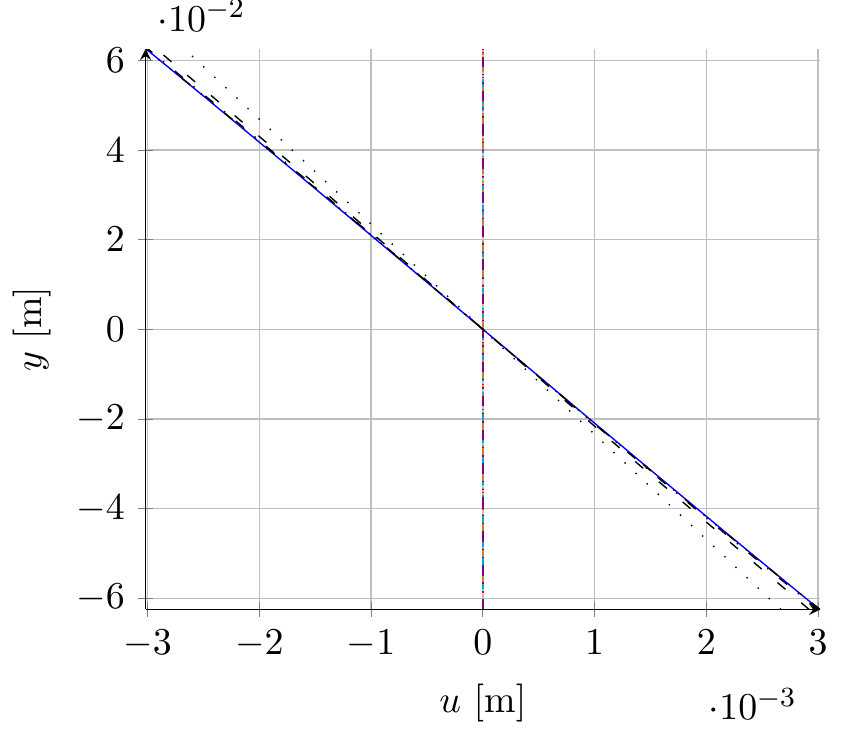}}
\hfil
\subfigure[$\nu = 0$]{\includegraphics[width=0.41\textwidth]{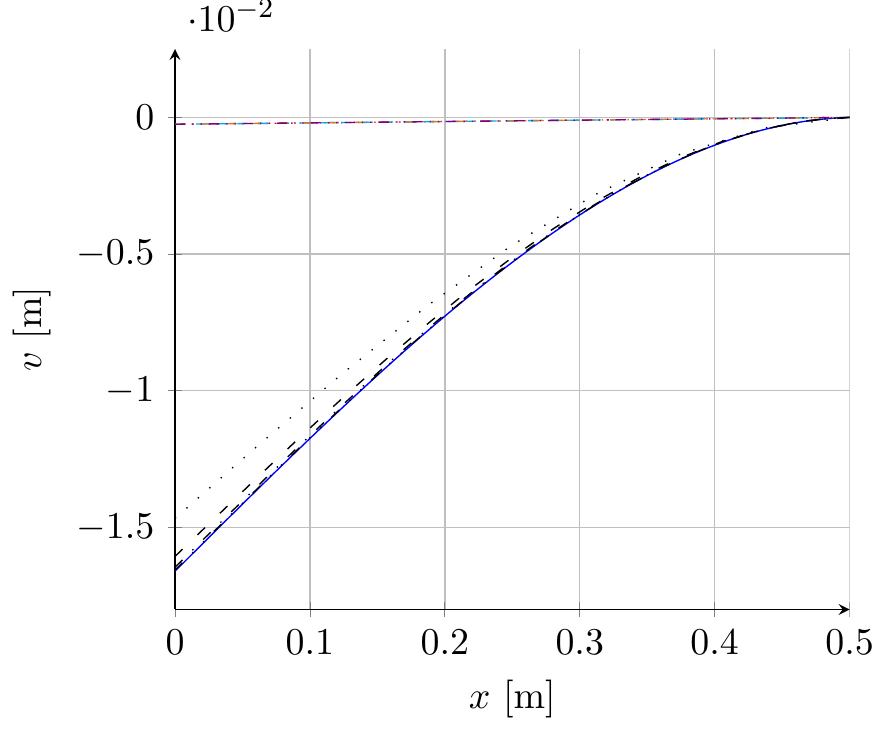}}
\vfill
\subfigure[$\nu = 0.3$]{\includegraphics[width=0.4\textwidth]{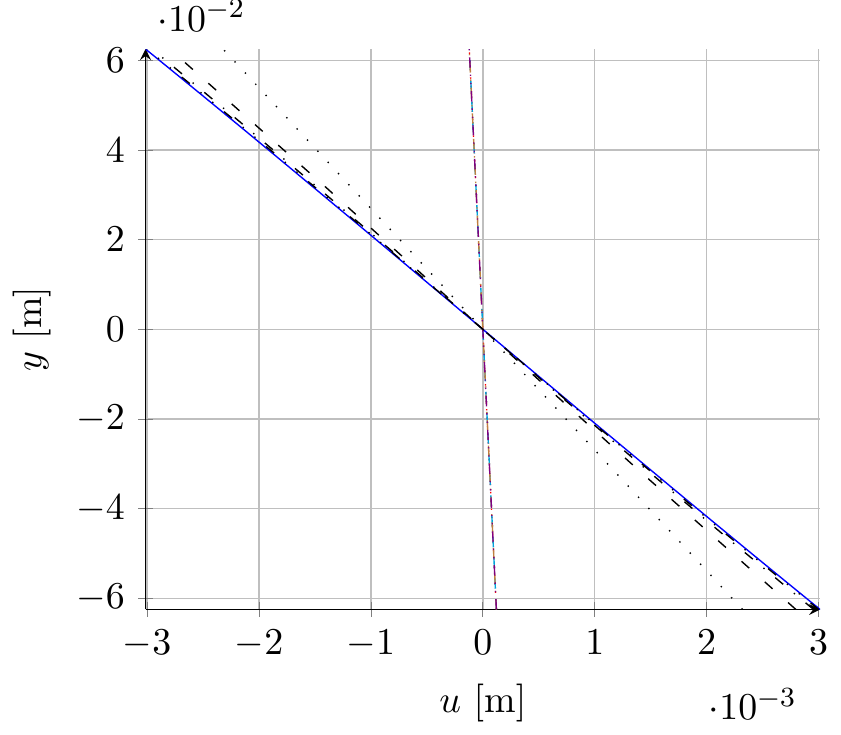}}
\hfil
\subfigure[$\nu = 0.3$]{\includegraphics[width=0.41\textwidth]{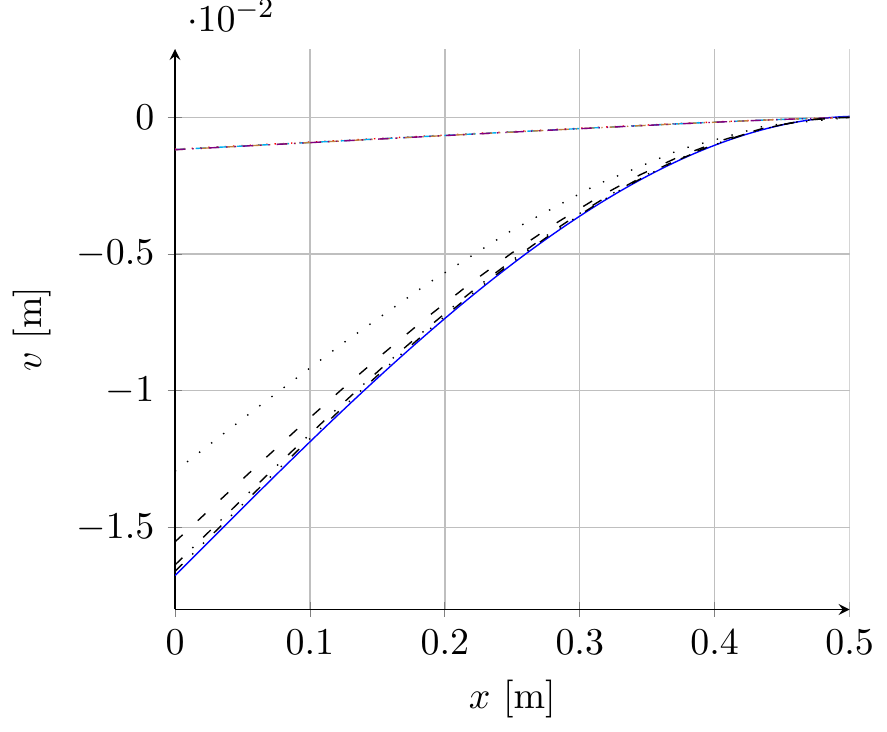}}
\vfill
\subfigure[$\nu = 0.49$]{\includegraphics[width=0.4\textwidth]{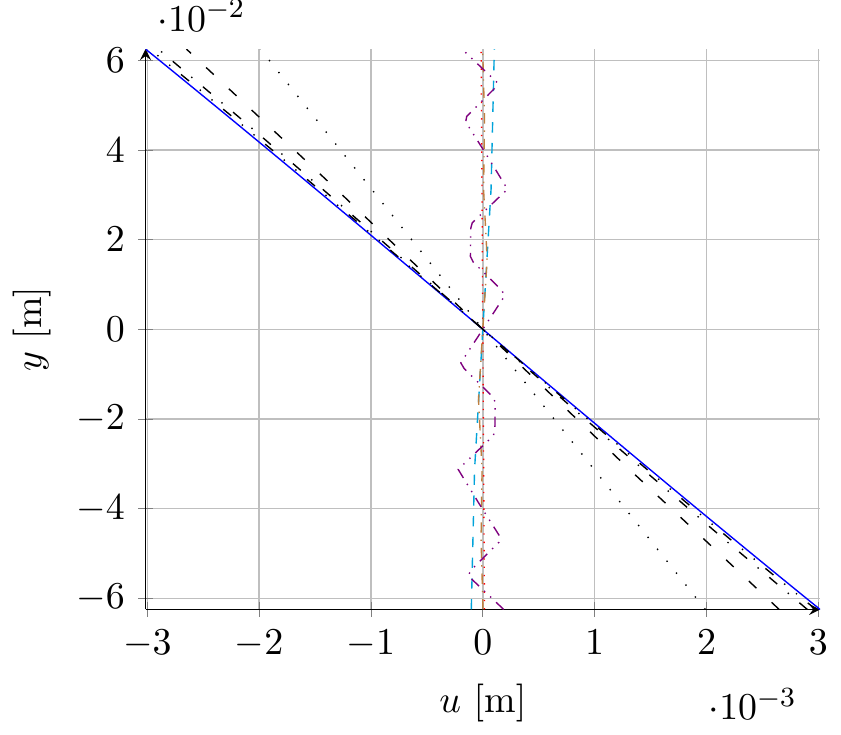}}
\hfil
\subfigure[$\nu = 0.49$]{\includegraphics[width=0.41\textwidth]{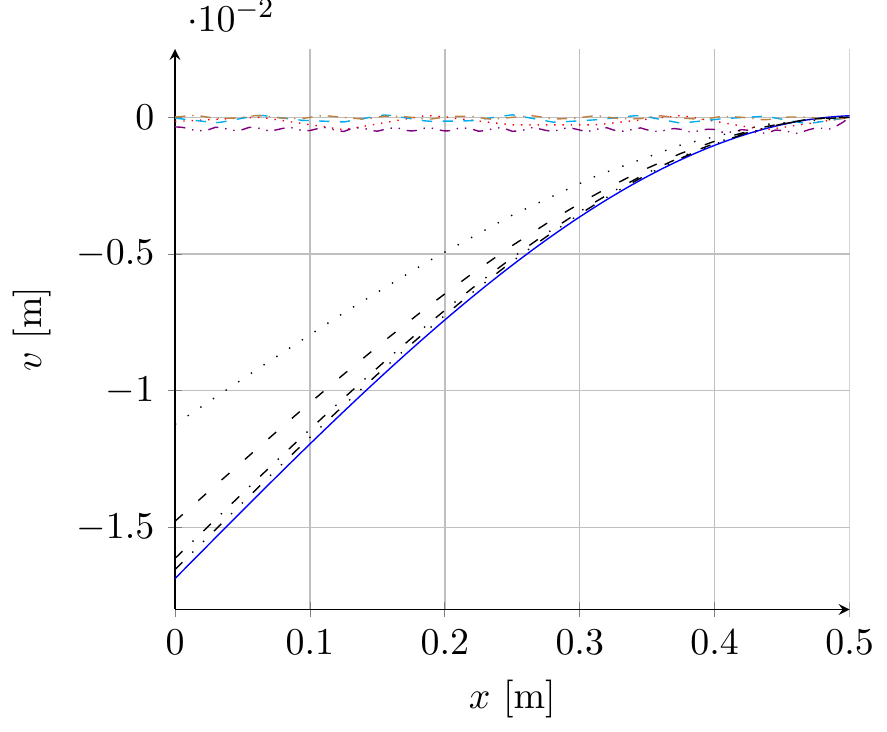}}
\caption{Displacements $u$ at the left edge of the plate (left) and $v$ of the axis of the plate (right) employing the \textit{Born} model and the modified model}
\label{fig: comparison of displacements cantilever bending}
\end{figure}

The distributions of the displacement $u$ at the left edge of the plate and the displacement $v$ of the axis of the plate regarding the \textit{Born model} and the modified model are summarised in Figure \ref{fig: comparison of displacements cantilever bending}. From the results for $\nu < 1/3$ we observe that the response of the plate discretised with the unit-cell employing the \textit{Born model} is stiffer in comparison with that of the analytical solution. This is because, for the case of cantilever bending there exists elements faraway from the boundary that undergo rigid-body rotation as well. Since the \textit{Born model} cannot distinguish between rigid-body rotations and shear deformations, strain energy is stored also for rigid-body rotation. For $\nu > 1/3$ the \textit{Born model} produces unstable results. The reason for this behaviour can be understood by looking at the variation of normalised eigenvalues of an unconstrained unit-cell stiffness matrix given in Figure \ref{fig: variation of eigenvalues of unit-cell}(a). In the range $0 < \nu < 1/3$, the eigenvalue $\lambda_{3}$ of the rigid body rotation eigenform is non-zero and therefore a part of the work performed by the external force is stored as strain energy due to this rigid-body motion. Only the remaining part of the work performed is available for the bending and shear eigenforms which are also included in the cantilever bending solution. For $\nu > 1/3$ the eigenvalue of the rigid-body rotation eigenform becomes negative and hence the strain energy function is no longer positive definite. As $\nu$ approaches the value $1/3$, $\lambda_{3}$ reduces to zero and this implies that the strain energy stored due to rigid-body rotation decreases. Therefore the response of the plate employing \textit{Born model} becomes relatively softer for $\nu = 0.3$ in comparison with the results for $\nu = 0$. Due to the presence of a non-zero eigenvalue for the rigid-body rotation eigenform, the solution also does not converge towards the analytical solution. Since the modified model is able to distinguish between rigid-body rotations and shear deformations, no strain energy is stored in the case of rigid-body rotation. Therefore the unit-cell employing the modified bond model produces satisfactory results. Also the solution converges towards the analytical solution upon refinement.

\newpage
\section{Conclusion}
\label{sec: conclusion}
Modelling of continuum isotropic elasticity with particle methods such as the discrete element method or the lattice spring method is traditionally limited to a certain value of Poisson's ratio due to the bond model used. The modified bond model presented in this paper for planar continuum overcomes this limitation by introducing a multi-bond term that couples the shear strain energy of neighbour bonds. This term enables the modified bond model to distinguish between rigid-body rotation and shear deformation. Two bonds, namely the \textit{L-bond} and \textit{X-bond} that employ this proposed coupling were introduced. The positive definiteness of the strain energy function of the unit-cell employing the modified bond model was ensured for values of Poisson's ratio in the range $0 \leq \nu < 0.5$. The results obtained under uniaxial, pure bending, cantilever bending and pure shear loadings were validated with the continuum mechanics solution. Moreover, the agreement of the numerical results with the continuum mechanics solution demonstrate the ability of the modified bond model to describe the behaviour of an isotropic elastic material. The concept of shear strain energy coupling of neighbour bonds provides an alternate method to those existing in the literature which are based on computation of a particle stress or strain tensor. The generalisation of this modified bond model for random discretisation and for three-dimensions are under progress. The implementation of this modified bond model within the framework of the DEM is also to be carried out.

\scriptsize
\appendix
\section{Stiffness matrix of the unit-cell with \textit{Born model}}
\label{appendix: stiffness matrix of the unit-cell with Born bond model}
In the work of Griffiths \cite{Griffiths:2001aa}, the stiffness matrix of a unit-cell was obtained by assembling the stiffness matrix of individual constituent bonds in a procedure that is similar to that of the Finite Element Method (FEM). Here an alternative approach of deriving the stiffness matrix from the strain energy of the unit-cell is taken. The strain energy stored in a generic bond in terms of the local displacement components was given in Equation (\ref{eqn: strain energy in bond in terms of local displacements}). The local displacements of particles in a generic bond oriented at an angle $\theta$ to the global coordinate system can be written in terms of the global displacement with the transformation matrix $\mathbf{Q}$ as $\mathbf{u}_{lo} = \mathbf{Q} \, \mathbf{u}_{gl}$,
\begin{align}
	\begin{bmatrix}
		u_{A}^{n}\\
		v_{A}^{s}
	\end{bmatrix} = %
	\begin{bmatrix}
		\phantom{+}\cos \theta & \phantom{+} \sin \theta\\
		-\sin \theta & \phantom{+} \cos \theta
	\end{bmatrix} \,%
	\begin{bmatrix}
		u_{A}\\
		v_{A}
	\end{bmatrix} \hspace{5mm} &\text{and} \hspace{5mm}
	\begin{bmatrix}
		u_{B}^{n}\\
		v_{B}^{s}
	\end{bmatrix} = %
	\begin{bmatrix}
		\phantom{+}\cos \theta & \phantom{+} \sin \theta\\
		-\sin \theta & \phantom{+} \cos \theta
	\end{bmatrix} \,%
	\begin{bmatrix}
		u_{B}\\
		v_{B}
	\end{bmatrix} \label{eqn: transformation of displacements expanded}.
\end{align}
With this transformation, the strain energy in a bond in terms of the global particle displacements is given by
\begin{align} \label{eqn: strain energy in bond in terms of global displacement}
\begin{aligned}
	\Pi_{b} &= \frac{1}{2} \, k_{n} (\cos \theta (u_{B} - u_{A}) + \sin \theta (v_{B} - v_{A}))^{2}\\ 
	&+ \frac{1}{2} \, k_{s} (\cos \theta (v_{B} - v_{A}) + \sin \theta (u_{A} - u_{B}))^{2}. 
\end{aligned}
\end{align}
The strain energy of the unit-cell in terms of displacements are obtained by summing up the strain energy of individual constituent bonds given in equation (\ref{eqn: strain energy in bond in terms of global displacement}) by taking in to account their orientation as 
\begin{align} \label{eqn: summation of strain energy in bond of unit-cell in terms of displacement}
\begin{aligned}
	\Pi_{uc} &= \Pi_{b}(\theta = 0^{\circ}) + \Pi_{b}(\theta = 90^{\circ}) + \Pi_{b}(\theta = 180^{\circ}) + \Pi_{b}(\theta = 270^{\circ})\\ 
	&+ \Pi_{b}(\theta = 45^{\circ})+ \Pi_{b}(\theta = 135^{\circ}).
\end{aligned}
\end{align} 
While summing up, the stiffness factor of individual bonds have to be considered as well. The stiffness factor for bonds with first neighbours is $1/2$, since they are shared by two unit-cells (periodicity). However, the stiffness factor for bonds with second neighbours is $1$ as they belong exclusively to each unit-cell. By using the appropriate stiffness factors and by substituting the orientation of individual bonds, the stiffness matrix of the unit-cell is obtained by differentiating equation (\ref{eqn: summation of strain energy in bond of unit-cell in terms of displacement}) twice with respect to the appropriate global displacement as

\begin{minipage}[H]{0.5\textwidth}
	\begin{align} 
	\mathbf{K}_{uc} = %
	\begin{bmatrix}
		\phantom{+}\hat{K}_{1} & \phantom{+0} & \phantom{+0} & \phantom{+0} & \phantom{+0} & \phantom{+0} & \phantom{+0} & \phantom{+0}\\
		\phantom{+}\hat{K}_{2} & \phantom{+}\hat{K}_{1} & \phantom{+0} & \phantom{+0} & \phantom{+0} & \phantom{+0} & \phantom{+0} & \phantom{+0}\\
		\phantom{+}\hat{K}_{3} & \phantom{+} 0 & \phantom{+}\hat{K}_{1} & \phantom{+0} & \phantom{+0} & \phantom{+0} & \phantom{+0} & \phantom{+0}\\
		\phantom{+}0 & \phantom{+}\hat{K}_{5} & -\hat{K}_{2} & \phantom{+}\hat{K}_{1} & \phantom{+0} & \phantom{+0} & \phantom{+0} & \phantom{+0}\\
		\phantom{+}\hat{K}_{4} & -\hat{K}_{2} & \phantom{+}\hat{K}_{5} & \phantom{+} 0 & \phantom{+}\hat{K}_{1} & \phantom{+0} & \phantom{+0} & \phantom{+0}\\
		-\hat{K}_{2} & \phantom{+}\hat{K}_{4} & \phantom{+} 0 & \phantom{+}\hat{K}_{3} & \phantom{+}\hat{K}_{2} & \phantom{+}\hat{K}_{1} & \phantom{+0} & \phantom{+0}\\
		\phantom{+}\hat{K}_{5} & \phantom{+} 0 & \phantom{+}\hat{K}_{4} & \phantom{+}\hat{K}_{2} & \phantom{+}\hat{K}_{3} & \phantom{+} 0 & \phantom{+}\hat{K}_{1} & \phantom{+0}\\
		\phantom{+} 0 & \phantom{+}\hat{K}_{3} & \phantom{+}\hat{K}_{2} & \phantom{+}\hat{K}_{4} & \phantom{+} 0 & \phantom{+}\hat{K}_{5} & -\hat{K}_{2} & \phantom{+}\hat{K}_{1}
	\end{bmatrix} \nonumber
\end{align}
\end{minipage}
\begin{minipage}[H]{0.4\textwidth}
	\begin{align} \label{eqn: stiffness matrix of unit-cell with Born model}
\begin{aligned}
	\hat{K}_{1} &= \phantom{+}\frac{1}{2}k_{n_{1}}^{b} + \frac{1}{2}k_{n_{2}}^{b} + k_{s_{1}}^{b}\\
	\hat{K}_{2} &= \phantom{+}\frac{1}{2}k_{n_{2}}^{b} - \frac{1}{2} k_{s_{1}}^{b}\\
	\hat{K}_{3} &= -\frac{1}{2}k_{n_{1}}^{b}\\
	\hat{K}_{4} &= -\frac{1}{2}k_{n_{2}}^{b} - \frac{1}{2} k_{s_{1}}^{b}\\
	\hat{K}_{5} &= -\frac{1}{2}k_{s_{1}}^{b}.
\end{aligned}
\end{align}
\end{minipage}

\section{Stiffness matrix of the unit-cell with the modified bond model}
\label{appendix: stiffness matrix of the unit-cell with improved bond model}
For a generic \textit{L-bond} as shown in Figure \ref{fig: first uc with l-bonds}(a), the strain energy stored in terms of the local displacements of the particles can be written as
\begin{align} \label{eqn: strain energy in generic l-bond in terms of local displacements}
	\Pi_{ABC} = \frac{1}{2} k_{n_{1}}^{m} \bigg[ (u_{B}^{n} - u_{A}^{n})^{2} + (u_{C}^{n} - u_{B}^{n})^{2} \bigg] + \frac{1}{2} k_{s_{1}}^{m} \bigg[ -(v_{B}^{s} - v_{A}^{s}) + (v_{C}^{s} - v_{B}^{s}) \bigg]^{2}.
\end{align}
With the transformation matrix $\mathbf{Q}$ and Equation (\ref{eqn: strain energy in bond in terms of global displacement}), the strain energy is now given in terms of the global particle displacements as
\begin{align}
\begin{aligned}
	\Pi_{ABC} &= \frac{1}{2} k_{n_{1}}^{m} \bigg[ \bigg( (c_{AB} \, u_{B} + s_{AB} \, v_{B}) - (c_{AB} \, u_{A} + s_{AB} \, v_{A}) \bigg)^{2}\\
	&+ \bigg( (c_{BC} \, u_{C} + s_{BC} \, v_{C}) - (c_{BC} \, u_{B} + s_{BC} \, v_{B}) \bigg)^{2} \bigg]\\
	&+ \frac{1}{2} k_{s_{1}}^{m} \bigg[ -s_{BC} \, (u_{C} - u_{B}) + c_{BC} \, (v_{C} - v_{B}) + s_{AB} \, (u_{B} - u_{A}) - c_{AB} \, (v_{B} - v_{A}) \bigg]^{2}
\end{aligned}	
\end{align}
where, $c_{AB} = \cos \theta_{AB}$, $s_{AB} = \sin \theta_{AB}$, $c_{BC} = \cos \theta_{BC}$ and $s_{BC} = \sin \theta_{BC}$. The strain energy stored in the unit-cell made up of \textit{L-bonds} $\Pi_{uc_{1}}^{mod}$ is obtained by summing up the strain energy of individual constituent \textit{L-bond} as shown in Figure \ref{fig: first uc with l-bonds}(b) and in Equation (\ref{eqn: strain energy in first neighbour unit-cell with improved bond model}). Similarly the strain energy stored in the unit-cell made up of \textit{X-bond} is given in terms of the local displacements as
\begin{align} \label{eqn: strain energy in generic x-bond in terms of local displacements}
	\Pi_{uc_{2}}^{mod} = \frac{1}{2} k_{n_{2}}^{m} \bigg[ (u_{C}^{n} - u_{A}^{n})^{2} + (u_{D}^{n} - u_{B}^{n})^{2} \bigg] + \frac{1}{2} k_{s_{1}}^{m} \bigg[ (v_{C}^{s} - v_{A}^{s}) - (v_{D}^{s} - v_{B}^{s}) \bigg]^{2}.
\end{align}
Similar to \textit{L-bond}, the local displacements are expressed in terms of the global displacements with the transformation matrix $\mathbf{Q}$. After this, the strain energy stored in the unit-cell with the modified bond model is given by
\begin{align}
	\Pi_{uc}^{mod} = \Pi_{uc_{1}}^{mod} + \Pi_{uc_{2}}^{mod}.
\end{align}
By differentiating the strain energy twice with respect to the appropriate global displacements, the stiffness matrix is given by\newline
\begin{minipage}[H]{0.5\textwidth}
	\begin{align}
		\mathbf{K}_{uc}^{mod} = %
	\begin{bmatrix}
		\phantom{+}\hat{K}_{1} & \phantom{+0} & \phantom{+0} & \phantom{+0} & \phantom{+0} & \phantom{+0} & \phantom{+0} & \phantom{+0}\\
		\phantom{+}\hat{K}_{2} & \phantom{+}\hat{K}_{1} & \phantom{+0} & \phantom{+0} & \phantom{+0} & \phantom{+0} & \phantom{+0} & \phantom{+0}\\
		\phantom{+}\hat{K}_{3} & -\hat{K}_{4} & \phantom{+}\hat{K}_{1} & \phantom{+0} & \phantom{+0} & \phantom{+0} & \phantom{+0} & \phantom{+0}\\
		\phantom{+}\hat{K}_{4} & \phantom{+}0 & -\hat{K}_{2} & \phantom{+}\hat{K}_{1} & \phantom{+0} & \phantom{+0} & \phantom{+0} & \phantom{+0}\\
		\phantom{+}\hat{K}_{5} & -\hat{K}_{2} & \phantom{+}0 & -\hat{K}_{4} & \phantom{+}\hat{K}_{1} & \phantom{+0} & \phantom{+0} & \phantom{+0}\\
		-\hat{K}_{2} & \phantom{+}\hat{K}_{5} & \phantom{+} \hat{K}_{4} & \phantom{+}\hat{K}_{3} & \phantom{+}\hat{K}_{2} & \phantom{+}\hat{K}_{1} & \phantom{+0} & \phantom{+0}\\
		\phantom{+}0 & \phantom{+} \hat{K}_{4} & \phantom{+}\hat{K}_{5} & \phantom{+}\hat{K}_{2} & \phantom{+}\hat{K}_{3} & -\hat{K}_{4} & \phantom{+}\hat{K}_{1} & \phantom{+0}\\
		-\hat{K}_{4} & \phantom{+}\hat{K}_{3} & \phantom{+}\hat{K}_{2} & \phantom{+}\hat{K}_{5} & \phantom{+} \hat{K}_{4} & \phantom{+}0 & -\hat{K}_{2} & \phantom{+}\hat{K}_{1}
	\end{bmatrix} \nonumber
	\end{align}
\end{minipage}
\begin{minipage}[H]{0.4\textwidth}
	\begin{align} \label{eqn: stiffness matrix of unit-cell with improved bond model}
\begin{aligned}
	\hat{K}_{1} &= \phantom{+}\frac{1}{2}k_{n_{1}}^{m} + \frac{1}{2}k_{n_{2}}^{m} + k_{s_{1}}^{m}\\
	\hat{K}_{2} &= \phantom{+}\frac{1}{2}k_{n_{2}}^{m} - \frac{1}{4} k_{s_{1}}^{m}\\
	\hat{K}_{3} &= -\frac{1}{2}k_{n_{1}}^{m} -\frac{1}{2}k_{s_{1}}^{m}\\
	\hat{K}_{4} &= -\frac{3}{4}k_{s_{1}}^{m}\\
	\hat{K}_{5} &= -\frac{1}{2}k_{n_{2}}^{m} - \frac{1}{2} k_{s_{1}}^{m}.
\end{aligned}
\end{align}
\end{minipage}

\bibliographystyle{spmpsci}      
\bibliography{references}

\end{document}

%% file: tikz/figure1a.tex
\begin{tikzpicture}
	
	\foreach \x in {0,1,...,10}
		\foreach \y in {0,1,...,5}
		{
			\draw [fill=gray!50,gray!50] (\x, \y) circle (0.4925);
			\draw [fill=black] (\x,\y) circle (0.05);
		}
	
	\draw [step=1] (0,0) grid (10,5);
	\draw (0,0) -- (5,5);
	\draw (1,0) -- (6,5);
	\draw (2,0) -- (7,5);
	\draw (3,0) -- (8,5);
	\draw (4,0) -- (9,5);
	\draw (5,0) -- (10,5);
	\draw (6,0) -- (10,4);
	\draw (7,0) -- (10,3);
	\draw (8,0) -- (10,2);
	\draw (9,0) -- (10,1);
	\draw (0,1) -- (4,5);
	\draw (0,2) -- (3,5);
	\draw (0,3) -- (2,5);
	\draw (0,4) -- (1,5);
	
	\draw (0,1) -- (1,0);
	\draw (0,2) -- (2,0);
	\draw (0,3) -- (3,0);
	\draw (0,4) -- (4,0);
	\draw (0,5) -- (5,0);
	\draw (1,5) -- (6,0);
	\draw (2,5) -- (7,0);
	\draw (3,5) -- (8,0);
	\draw (4,5) -- (9,0);
	\draw (5,5) -- (10,0);
	\draw (6,5) -- (10,1);
	\draw (7,5) -- (10,2);
	\draw (8,5) -- (10,3);
	\draw (9,5) -- (10,4);
\end{tikzpicture}	

%% file: tikz/figure1b.tex
\begin{tikzpicture}[scale=2]
		
	\draw (0,0) rectangle (1,1);
	\draw (0,0) -- (1,1);
	\draw (1,0) -- (0,1);
	
	\draw [fill] (0,0) circle (0.05) node[left] {$A$};
	\draw [fill] (1,0) circle (0.05) node[right] {$B$};
	\draw [fill] (1,1) circle (0.05) node[right] {$C$};	
	\draw [fill] (0,1) circle (0.05) node[left] {$D$};	
	
\end{tikzpicture}

%% file: tikz/figure1c.tex
\begin{tikzpicture}[scale=2]
		
\draw (0,0) rectangle (1,1);

\draw [fill] (0,0) circle (0.05) node[left] {$A$};
\draw [fill] (1,0) circle (0.05) node[right] {$B$};
\draw [fill] (1,1) circle (0.05) node[right] {$C$};
\draw [fill] (0,1) circle (0.05) node[left] {$D$};
			
\end{tikzpicture}

%% file: tikz/figure1d.tex
\begin{tikzpicture}[scale=2]

\draw (0, 0) -- (1, 1);
\draw (1, 0) -- (0, 1);

\draw [fill] (0, 0) circle (0.05) node[left] {$A$};
\draw [fill] (1, 0) circle (0.05) node[right] {$B$};
\draw [fill] (1, 1) circle (0.05) node[right] {$C$};
\draw [fill] (0, 1) circle (0.05) node[left] {$D$};

\end{tikzpicture}

%% file: tikz/figure2a.tex
\begin{tikzpicture}[scale=1.5, line cap=round]
\tikzstyle{spring_x}=[thin,decorate,decoration={zigzag,pre length=0.5cm,post length=0.5cm,segment length=8}]

\tikzstyle{spring_y}=[thin,decorate,decoration={zigzag,pre length=0.25cm,post length=0.25cm,segment length=4}]

\draw [spring_x] (0,0) -- (2,0);
\draw [spring_y] (1,0.5) -- (1,-0.5);

\draw (0,0) -- (1,0.5);
\draw (2,0) -- (1,-0.5);

\draw [fill=black] (0,0) circle (0.05) node[left] {$A$};
\draw [fill=black] (2,0) circle (0.05) node[right] {$B$};

\node[above] at (1.75, 0) {$k_{n}$};
\node[right] at (1, 0.45) {$k_{s}$};
	
\end{tikzpicture}

%% file: tikz/figure2b.tex
\begin{tikzpicture}[scale=2, line cap=round]
	
\draw (0,0) -- (2,1);
\draw [fill=black] (0,0) circle (0.05) node[left] {$A$};
\draw [fill=black] (2,1) circle (0.05) node[right] {$B$};

\draw [dashed] (0,0) -- (1,0);

\draw [->,>=latex,thin] (-1,0) -- (-0.5,0) node[below] {$x, \, u$};
\draw [->,>=latex,thin] (-1,0) -- (-1,0.5) node[above] {$y, \, v$};

\draw [->,>=latex] (0.9,0.7) -- (1.3,0.9) node[right] {$n, \, u_{n}$};
\begin{scope}[rotate around={90:(0.9,0.7)}]
\draw [->,>=latex] (0.9,0.7) -- (1.3,0.9) node[above] {$s, \, v_{s}$};			
\end{scope}

\begin{scope}[shift={(-0.5,1)}]
\draw [|<->|, >=latex] (0,0) -- (2,1) node[midway, above] {$l$};
\end{scope}

\draw [very thin, ->, >=latex] (0.7,0) arc [start angle=0, end angle=26, radius=0.7] node[midway, right] {$\theta$};

\end{tikzpicture}

%% file: tikz/figure3a.tex
\begin{tikzpicture}[scale=0.8, line cap=round]
	
	\draw [gray] (4,2) -- (6,2);
	\draw [fill=gray, gray] (4,2) circle (0.05) node[left, black] {$A$};
	\draw [fill=gray, gray] (6,2) circle (0.05) node[right, black] {$B$};
		
	\draw (4,1.75) -- (6,2.25);
	\draw [fill=black] (4,1.75) circle (0.05);
	\draw [fill=black] (6,2.25) circle (0.05);
		
	\draw [->, >=latex] (5, 2.5) -- (5.5, 2.5) node[right] {$n$};
	\draw [->, >=latex] (5, 2.5) -- (5, 3) node[above] {$s$};
		
	\node at (5, 1.5) {$\epsilon_{sn}^{AB} = \epsilon_{yx}$};
		
	\draw [gray] (7.5,3.5) -- (7.5,5.5);
	\draw [fill=gray, gray] (7.5,3.5) circle (0.05) node[below, black] {$B$};
	\draw [fill=gray, gray] (7.5,5.5) circle (0.05) node[above, black] {$C$};
		
	\draw (7.75, 3.5) -- (7.25, 5.5);
	\draw [fill=black] (7.75, 3.5) circle (0.05);
	\draw [fill=black] (7.25, 5.5) circle (0.05);
		
	\draw [->, >=latex] (7, 4.5) -- (7, 5) node[above] {$n$};
	\draw [->, >=latex] (7, 4.5) -- (6.5, 4.5) node[left] {$s$};

	\node [rotate=90] at (8, 4.5) {$\epsilon_{sn}^{BC} = -\epsilon_{xy}$};
		
	\draw [gray] (6, 7) -- (4, 7);
	\draw [fill=gray, gray] (6, 7) circle (0.05) node[right, black] {$C$};
	\draw [fill=gray, gray] (4, 7) circle (0.05) node[left, black] {$D$};
		
	\draw (6, 7.25) -- (4, 6.75);
	\draw [fill=black] (6, 7.25) circle (0.05);
	\draw [fill=black] (4, 6.75) circle (0.05);
		
	\draw [->, >=latex] (5, 6.5) -- (4.5, 6.5) node[left] {$n$};
	\draw [->, >=latex] (5, 6.5) -- (5, 6) node[below] {$s$};
		
	\node at (5, 7.5) {$\epsilon_{sn}^{CD} = \epsilon_{yx}$};
		
	\draw [gray] (2.5, 5.5) -- (2.5, 3.5);
	\draw [fill=gray, gray] (2.5, 5.5) circle (0.05) node[above, black] {$D$};
	\draw [fill=gray, gray] (2.5, 3.5) circle (0.05) node[below, black] {$A$};
		
	\draw (2.25, 5.5) -- (2.75, 3.5);
	\draw [fill=black] (2.25, 5.5) circle (0.05);
	\draw [fill=black] (2.75, 3.5) circle (0.05);
		
	\draw [->, >=latex] (3, 4.5) -- (3, 4) node[below] {$n$};
	\draw [->, >=latex] (3, 4.5) -- (3.5, 4.5) node[right] {$s$};
		
	\node [rotate=90] at (2, 4.5) {$\epsilon_{sn}^{DA} = -\epsilon_{xy}$};
		
\end{tikzpicture}

%% file: tikz/figure3b.tex
\begin{tikzpicture}[scale=1.5, line cap=round]
		
	\draw [gray] (2,2) -- (4,2) -- (4,4) -- (2,4) -- cycle;
	\draw [fill=gray, gray] (2,2) circle (0.05) node[left, black] {$A$};
	\draw [fill=gray, gray] (4,2) circle (0.05) node[right, black] {$B$};
	\draw [fill=gray, gray] (4,4) circle (0.05) node[right, black] {$C$};
	\draw [fill=gray, gray] (2,4) circle (0.05) node[left, black] {$D$};
		
	\coordinate (A) at (2,2) {};
	\coordinate (C) at (4,4) {};
	\coordinate (D) at (2,4) {};
		
	\tkzMarkRightAngle[very thin, draw=gray](C,D,A);

	\draw (2.2,1.8) -- (4.2,2.2) -- (3.8,4.2) -- (1.8,3.8) -- cycle;
	\draw [fill=black] (2.2,1.8) circle (0.05);
	\draw [fill=black] (4.2,2.2) circle (0.05);
	\draw [fill=black] (3.8,4.2) circle (0.05);
	\draw [fill=black] (1.8,3.8) circle (0.05);
		
	\coordinate (A_new) at (2.2,1.8) {};
	\coordinate (C_new) at (3.8,4.2) {};
	\coordinate (D_new) at (1.8,3.8) {};
		
	\tkzMarkRightAngle[very thin](C_new,D_new,A_new);
		
\end{tikzpicture}

%% file: tikz/figure3c.tex
\begin{tikzpicture}[scale=1.1, line cap=round]
	
	\draw[gray] (1, 1) -- (3, 3);
	\draw[fill=gray, gray] (1, 1) circle (0.05) node[below, black] {$A$};
	\draw[fill=gray, gray] (3, 3) circle (0.05) node[above, black] {$C$};
	
	\draw (1.25, 0.75) -- (2.75, 3.25);
	\draw[fill] (1.25, 0.75) circle (0.05);
	\draw[fill] (2.75, 3.25) circle (0.05);
	
	\draw[->, >=latex] (1.5, 2.5) -- (2, 3) node[above] {$n$};
	\draw[->, >=latex] (1.5, 2.5) -- (1, 3) node[above] {$s$};
	
	\draw[gray] (7, 1) -- (5, 3);
	\draw[fill=gray, gray] (7, 1) circle (0.05) node[below, black] {$B$};
	\draw[fill=gray, gray] (5, 3) circle (0.05) node[above, black] {$D$};
	
	\draw (7.25, 1.25) -- (4.75, 2.75);
	\draw[fill] (7.25, 1.25) circle (0.05);
	\draw[fill] (4.75, 2.75) circle (0.05);
	
	\draw[->, >=latex] (5.5, 1.5) -- (6, 1) node[below] {$n$};
	\draw[->, >=latex] (5.5, 1.5) -- (5, 1) node[below] {$s$};
		
\end{tikzpicture}

%% file: tikz/figure3d.tex
\begin{tikzpicture}[scale=1.1, line cap=round]
	
	\draw[gray] (1, 1) -- (3, 3);
	\draw[gray] (3, 1) -- (1, 3);
	
	\draw[fill=gray, gray] (1, 1) circle (0.05) node[below, black] {$A$};
	\draw[fill=gray, gray] (3, 1) circle (0.05) node[below, black] {$B$};
	\draw[fill=gray, gray] (3, 3) circle (0.05) node[above, black] {$C$};
	\draw[fill=gray, gray] (1, 3) circle (0.05) node[above, black] {$D$};
	
	\draw (1.25, 0.75) -- (2.75, 3.25);
	\draw (3.25, 1.25) -- (0.75, 2.75);
	
	\draw[fill] (1.25, 0.75) circle (0.05);
	\draw[fill] (2.75, 3.25) circle (0.05);
	\draw[fill] (3.25, 1.25) circle (0.05);
	\draw[fill] (0.75, 2.75) circle (0.05);
	
	\coordinate (A) at (1, 1) {};
	\coordinate (B) at (2, 2) {};
	\coordinate (C) at (3, 1) {};
	
	\tkzMarkRightAngle[very thin, draw=gray, size=.15](A,B,C);
	
	\coordinate (A_new) at (2.75, 3.25) {};
	\coordinate (B_new) at (2, 2) {};
	\coordinate (C_new) at (0.75, 2.75) {};
	
	\tkzMarkRightAngle[very thin, draw=black, size=.15](A_new,B_new,C_new);
	
\end{tikzpicture}

%% file: tikz/figure4a.tex
\begin{tikzpicture}[scale=1.5, line cap=round]
		
	\draw (2, 4) -- (2, 2) -- (4, 2);
	\draw [fill=black] (2, 4) circle (0.05) node[above] {$A$};
	\draw [fill=black] (2, 2) circle (0.05) node[below] {$B$};
	\draw [fill=black] (4, 2) circle (0.05) node[below] {$C$};
		
	\draw [->,>=latex] (3, 1) -- (3.5, 1) node[right] {$n$};
	\draw [->,>=latex] (3, 1) -- (3, 1.5) node[left] {$s$};
		
	\draw [->,>=latex] (1, 3) -- (1, 2.5) node[below] {$n$};
	\draw [->,>=latex] (1, 3) -- (1.5, 3) node[above] {$s$};
		
	\draw [->,>=latex] (0.5, 0.5) -- (1, 0.5) node[right] {$x$};
	\draw [->,>=latex] (0.5, 0.5) -- (0.5, 1) node[above] {$y$};
		
\end{tikzpicture}

%% file: tikz/figure4b.tex
\begin{tikzpicture}[scale=0.9, line cap=round]

	\draw (1, 7) -- (1, 5) -- (3, 5);
	\draw (1, 3) -- (3, 3) -- (3, 1);
	\draw (4, 5) -- (6, 5) -- (6, 7);
	\draw (6, 3) -- (4, 3) -- (4, 1);
	
	\draw[fill] (1, 7) circle (0.05) node[above] {$D$};
	\draw[fill] (1, 5) circle (0.05) node[below] {$A$};
	\draw[fill] (3, 5) circle (0.05) node[below] {$B$};
	
	\draw[fill] (1, 3) circle (0.05) node[above] {$D$};
	\draw[fill] (3, 3) circle (0.05) node[above] {$C$};
	\draw[fill] (3, 1) circle (0.05) node[below] {$B$};
	
	\draw[fill] (4, 5) circle (0.05) node[below] {$A$};
	\draw[fill] (6, 5) circle (0.05) node[below] {$B$};
	\draw[fill] (6, 7) circle (0.05) node[above] {$C$};
	
	\draw[fill] (6, 3) circle (0.05) node[above] {$C$};
	\draw[fill] (4, 3) circle (0.05) node[above] {$D$};
	\draw[fill] (4, 1) circle (0.05) node[below] {$A$};
		
\end{tikzpicture}

%% file: tikz/figure7.tex
\begin{tikzpicture}[scale=0.6, line cap=round]

	\scaling{0.6};
	
	\normalsize

	\draw[thick] (1, 1) rectangle (7, 7);
		
	\point{a}{1}{1};
	\support{1}{a};
		
	\point{b}{3}{1};
	\support{2}{b};
		
	\point{c}{5}{1};
	\support{2}{c};
		
	\point{d}{7}{1};
	\support{2}{d};
		
	\point{g}{1}{3};
	\support{2}{g}[-90];
		
	\point{h}{1}{5};
	\support{2}{h}[-90];
		
	\point{i}{1}{7};
	\support{2}{i}[-90];
		
	\point{l}{9.25}{1};
	\point{m}{9.25}{7}
		
	\lineload{1}{l}{m};
	\node at (9.75, 4) {$\sigma_{xx}$};
		
	\dimensioning{1}{a}{d}{-1}[$l$];
		
	\draw[->, thick, >=latex] (1.5, 1.5) -- (2.5, 1.5) node[right] {$x$};
	\draw[->, thick, >=latex] (1.5, 1.5) -- (1.5, 2.5) node[above] {$y$};
		
\end{tikzpicture}

%% file: tikz/figure9.tex
\begin{tikzpicture}[scale=0.38, line cap=round]
	
		\scaling{0.38};
		
		\normalsize
		
		\draw [thick] (6,6) rectangle (16,16);
		
		\point{a}{6}{6};
		\support{1}{a};
		
		\point{b}{8}{6};
		\support{1}{b};
		
		\point{c}{10}{6};
		\support{1}{c};
		
		\point{d}{12}{6};
		\support{1}{d};
		
		\point{e}{14}{6};
		\support{1}{e};
		
		\point{f}{16}{6};
		\support{1}{f};
		
		\point{k}{6}{16};
		
		\point{l}{17}{6};
		\point{m}{17}{16}
		
		\lineload{4}{l}{m};
		
		\point{n}{6}{16.75};
		\point{o}{16}{16.75}
		
		\lineload{4}{n}{o};
		
		\point{p}{5}{15.75};
		\point{q}{5}{5.75}
		
		\lineload{4}{p}{q};
		
		
		\point{r}{16}{16};
		\dimensioning{2}{a}{k}{1}[$l$];
		
		\draw[->, thick, >=latex] (6.5,6.5) -- (7.5,6.5) node[right] {$x$};
		\draw[->, thick, >=latex] (6.5,6.5) -- (6.5,7.5) node[above] {$y$};
		
		\node at (19, 11) {$\sigma_{xy}$};
		\node at (3, 11) {$\sigma_{xy}$};
		\node at (11, 18.25) {$\sigma_{yx}$};
		
\end{tikzpicture}

%% file: figures/no_mod_nu0.pdf_tex
\begingroup%
  \makeatletter%
  \providecommand\color[2][]{%
    \errmessage{(Inkscape) Color is used for the text in Inkscape, but the package 'color.sty' is not loaded}%
    \renewcommand\color[2][]{}%
  }%
  \providecommand\transparent[1]{%
    \errmessage{(Inkscape) Transparency is used (non-zero) for the text in Inkscape, but the package 'transparent.sty' is not loaded}%
    \renewcommand\transparent[1]{}%
  }%
  \providecommand\rotatebox[2]{#2}%
  \newcommand*\fsize{\dimexpr\f@size pt\relax}%
  \newcommand*\lineheight[1]{\fontsize{\fsize}{#1\fsize}\selectfont}%
  \ifx\svgwidth\undefined%
    \setlength{\unitlength}{778.21168358bp}%
    \ifx\svgscale\undefined%
      \relax%
    \else%
      \setlength{\unitlength}{\unitlength * \real{\svgscale}}%
    \fi%
  \else%
    \setlength{\unitlength}{\svgwidth}%
  \fi%
  \global\let\svgwidth\undefined%
  \global\let\svgscale\undefined%
  \makeatother%
  \begin{picture}(1,0.3165344)%
    \lineheight{1}%
    \setlength\tabcolsep{0pt}%
    \put(0,0){\includegraphics[width=\unitlength,page=1]{./figures/no_mod_nu0.pdf}}%
    \put(0.42820403,0.23024087){\color[rgb]{0,0,0}\makebox(0,0)[lt]{\lineheight{1.25}\smash{\begin{tabular}[t]{l}$u$ [m]\end{tabular}}}}%
    \put(0.45760958,0.20439963){\color[rgb]{0,0,0}\makebox(0,0)[lt]{\lineheight{1.25}\smash{\begin{tabular}[t]{l}$1e-4$\end{tabular}}}}%
    \put(0.45760958,0.07430241){\color[rgb]{0,0,0}\makebox(0,0)[lt]{\lineheight{1.25}\smash{\begin{tabular}[t]{l}$0$\end{tabular}}}}%
    \put(0,0){\includegraphics[width=\unitlength,page=2]{./figures/no_mod_nu0.pdf}}%
    \put(0.0815751,0.00301624){\color[rgb]{0,0,0}\makebox(0,0)[lt]{\lineheight{1.25}\smash{\begin{tabular}[t]{l}$x$\end{tabular}}}}%
    \put(-0.00129506,0.08143103){\color[rgb]{0,0,0}\makebox(0,0)[lt]{\lineheight{1.25}\smash{\begin{tabular}[t]{l}$y$\end{tabular}}}}%
    \put(0,0){\includegraphics[width=\unitlength,page=3]{./figures/no_mod_nu0.pdf}}%
    \put(0.88533772,0.22862974){\color[rgb]{0,0,0}\makebox(0,0)[lt]{\lineheight{1.25}\smash{\begin{tabular}[t]{l}$v$ [m]\end{tabular}}}}%
    \put(0.91474327,0.2027885){\color[rgb]{0,0,0}\makebox(0,0)[lt]{\lineheight{1.25}\smash{\begin{tabular}[t]{l}$4.6e-5$\end{tabular}}}}%
    \put(0.91474327,0.07269127){\color[rgb]{0,0,0}\makebox(0,0)[lt]{\lineheight{1.25}\smash{\begin{tabular}[t]{l}$-4.6e-5$\end{tabular}}}}%
  \end{picture}%
\endgroup%

%% file: figures/no_mod_nu03.pdf_tex
\begingroup%
  \makeatletter%
  \providecommand\color[2][]{%
    \errmessage{(Inkscape) Color is used for the text in Inkscape, but the package 'color.sty' is not loaded}%
    \renewcommand\color[2][]{}%
  }%
  \providecommand\transparent[1]{%
    \errmessage{(Inkscape) Transparency is used (non-zero) for the text in Inkscape, but the package 'transparent.sty' is not loaded}%
    \renewcommand\transparent[1]{}%
  }%
  \providecommand\rotatebox[2]{#2}%
  \newcommand*\fsize{\dimexpr\f@size pt\relax}%
  \newcommand*\lineheight[1]{\fontsize{\fsize}{#1\fsize}\selectfont}%
  \ifx\svgwidth\undefined%
    \setlength{\unitlength}{778.21168358bp}%
    \ifx\svgscale\undefined%
      \relax%
    \else%
      \setlength{\unitlength}{\unitlength * \real{\svgscale}}%
    \fi%
  \else%
    \setlength{\unitlength}{\svgwidth}%
  \fi%
  \global\let\svgwidth\undefined%
  \global\let\svgscale\undefined%
  \makeatother%
  \begin{picture}(1,0.3165344)%
    \lineheight{1}%
    \setlength\tabcolsep{0pt}%
    \put(0,0){\includegraphics[width=\unitlength,page=1]{./figures/no_mod_nu03.pdf}}%
    \put(0.42820403,0.23024087){\color[rgb]{0,0,0}\makebox(0,0)[lt]{\lineheight{1.25}\smash{\begin{tabular}[t]{l}$u$ [m]\end{tabular}}}}%
    \put(0.45760958,0.20439963){\color[rgb]{0,0,0}\makebox(0,0)[lt]{\lineheight{1.25}\smash{\begin{tabular}[t]{l}$1.8e-4$\end{tabular}}}}%
    \put(0.45760958,0.07430241){\color[rgb]{0,0,0}\makebox(0,0)[lt]{\lineheight{1.25}\smash{\begin{tabular}[t]{l}$0$\end{tabular}}}}%
    \put(0,0){\includegraphics[width=\unitlength,page=2]{./figures/no_mod_nu03.pdf}}%
    \put(0.0815751,0.00301624){\color[rgb]{0,0,0}\makebox(0,0)[lt]{\lineheight{1.25}\smash{\begin{tabular}[t]{l}$x$\end{tabular}}}}%
    \put(-0.00129506,0.08143103){\color[rgb]{0,0,0}\makebox(0,0)[lt]{\lineheight{1.25}\smash{\begin{tabular}[t]{l}$y$\end{tabular}}}}%
    \put(0,0){\includegraphics[width=\unitlength,page=3]{./figures/no_mod_nu03.pdf}}%
    \put(0.88533772,0.22862974){\color[rgb]{0,0,0}\makebox(0,0)[lt]{\lineheight{1.25}\smash{\begin{tabular}[t]{l}$v$ [m]\end{tabular}}}}%
    \put(0.91474327,0.2027885){\color[rgb]{0,0,0}\makebox(0,0)[lt]{\lineheight{1.25}\smash{\begin{tabular}[t]{l}$3.8e-5$\end{tabular}}}}%
    \put(0.91474327,0.07269127){\color[rgb]{0,0,0}\makebox(0,0)[lt]{\lineheight{1.25}\smash{\begin{tabular}[t]{l}$-3.8e-5$\end{tabular}}}}%
    \put(0,0){\includegraphics[width=\unitlength,page=4]{./figures/no_mod_nu03.pdf}}%
  \end{picture}%
\endgroup%

%% file: figures/no_mod_nu049.pdf_tex
\begingroup%
  \makeatletter%
  \providecommand\color[2][]{%
    \errmessage{(Inkscape) Color is used for the text in Inkscape, but the package 'color.sty' is not loaded}%
    \renewcommand\color[2][]{}%
  }%
  \providecommand\transparent[1]{%
    \errmessage{(Inkscape) Transparency is used (non-zero) for the text in Inkscape, but the package 'transparent.sty' is not loaded}%
    \renewcommand\transparent[1]{}%
  }%
  \providecommand\rotatebox[2]{#2}%
  \newcommand*\fsize{\dimexpr\f@size pt\relax}%
  \newcommand*\lineheight[1]{\fontsize{\fsize}{#1\fsize}\selectfont}%
  \ifx\svgwidth\undefined%
    \setlength{\unitlength}{778.21168358bp}%
    \ifx\svgscale\undefined%
      \relax%
    \else%
      \setlength{\unitlength}{\unitlength * \real{\svgscale}}%
    \fi%
  \else%
    \setlength{\unitlength}{\svgwidth}%
  \fi%
  \global\let\svgwidth\undefined%
  \global\let\svgscale\undefined%
  \makeatother%
  \begin{picture}(1,0.31889711)%
    \lineheight{1}%
    \setlength\tabcolsep{0pt}%
    \put(0,0){\includegraphics[width=\unitlength,page=1]{./figures/no_mod_nu049.pdf}}%
    \put(0.42820403,0.23024087){\color[rgb]{0,0,0}\makebox(0,0)[lt]{\lineheight{1.25}\smash{\begin{tabular}[t]{l}$u$ [m]\end{tabular}}}}%
    \put(0.45760958,0.20439963){\color[rgb]{0,0,0}\makebox(0,0)[lt]{\lineheight{1.25}\smash{\begin{tabular}[t]{l}$3e-4$\end{tabular}}}}%
    \put(0.45760958,0.07430241){\color[rgb]{0,0,0}\makebox(0,0)[lt]{\lineheight{1.25}\smash{\begin{tabular}[t]{l}$-9.6e-5$\end{tabular}}}}%
    \put(0,0){\includegraphics[width=\unitlength,page=2]{./figures/no_mod_nu049.pdf}}%
    \put(0.0815751,0.00301624){\color[rgb]{0,0,0}\makebox(0,0)[lt]{\lineheight{1.25}\smash{\begin{tabular}[t]{l}$x$\end{tabular}}}}%
    \put(-0.00129506,0.08143103){\color[rgb]{0,0,0}\makebox(0,0)[lt]{\lineheight{1.25}\smash{\begin{tabular}[t]{l}$y$\end{tabular}}}}%
    \put(0,0){\includegraphics[width=\unitlength,page=3]{./figures/no_mod_nu049.pdf}}%
    \put(0.88533772,0.22862974){\color[rgb]{0,0,0}\makebox(0,0)[lt]{\lineheight{1.25}\smash{\begin{tabular}[t]{l}$v$ [m]\end{tabular}}}}%
    \put(0.91474327,0.2027885){\color[rgb]{0,0,0}\makebox(0,0)[lt]{\lineheight{1.25}\smash{\begin{tabular}[t]{l}$1.5e-5$\end{tabular}}}}%
    \put(0.91474327,0.07269127){\color[rgb]{0,0,0}\makebox(0,0)[lt]{\lineheight{1.25}\smash{\begin{tabular}[t]{l}$-1.5e-5$\end{tabular}}}}%
    \put(0,0){\includegraphics[width=\unitlength,page=4]{./figures/no_mod_nu049.pdf}}%
  \end{picture}%
\endgroup%

%% file: figures/shear_mod_nu049.pdf_tex
\begingroup%
  \makeatletter%
  \providecommand\color[2][]{%
    \errmessage{(Inkscape) Color is used for the text in Inkscape, but the package 'color.sty' is not loaded}%
    \renewcommand\color[2][]{}%
  }%
  \providecommand\transparent[1]{%
    \errmessage{(Inkscape) Transparency is used (non-zero) for the text in Inkscape, but the package 'transparent.sty' is not loaded}%
    \renewcommand\transparent[1]{}%
  }%
  \providecommand\rotatebox[2]{#2}%
  \newcommand*\fsize{\dimexpr\f@size pt\relax}%
  \newcommand*\lineheight[1]{\fontsize{\fsize}{#1\fsize}\selectfont}%
  \ifx\svgwidth\undefined%
    \setlength{\unitlength}{773.37409837bp}%
    \ifx\svgscale\undefined%
      \relax%
    \else%
      \setlength{\unitlength}{\unitlength * \real{\svgscale}}%
    \fi%
  \else%
    \setlength{\unitlength}{\svgwidth}%
  \fi%
  \global\let\svgwidth\undefined%
  \global\let\svgscale\undefined%
  \makeatother%
  \begin{picture}(1,0.31895231)%
    \lineheight{1}%
    \setlength\tabcolsep{0pt}%
    \put(0,0){\includegraphics[width=\unitlength,page=1]{./figures/shear_mod_nu049.pdf}}%
    \put(0.43088252,0.23168107){\color[rgb]{0,0,0}\makebox(0,0)[lt]{\lineheight{1.25}\smash{\begin{tabular}[t]{l}$u$ [m]\end{tabular}}}}%
    \put(0.46047201,0.20567819){\color[rgb]{0,0,0}\makebox(0,0)[lt]{\lineheight{1.25}\smash{\begin{tabular}[t]{l}$2.98e-4$\end{tabular}}}}%
    \put(0.46047201,0.07476718){\color[rgb]{0,0,0}\makebox(0,0)[lt]{\lineheight{1.25}\smash{\begin{tabular}[t]{l}$0$\end{tabular}}}}%
    \put(0,0){\includegraphics[width=\unitlength,page=2]{./figures/shear_mod_nu049.pdf}}%
    \put(0.08208536,0.0030351){\color[rgb]{0,0,0}\makebox(0,0)[lt]{\lineheight{1.25}\smash{\begin{tabular}[t]{l}$x$\end{tabular}}}}%
    \put(-0.00130316,0.08194039){\color[rgb]{0,0,0}\makebox(0,0)[lt]{\lineheight{1.25}\smash{\begin{tabular}[t]{l}$y$\end{tabular}}}}%
    \put(0,0){\includegraphics[width=\unitlength,page=3]{./figures/shear_mod_nu049.pdf}}%
    \put(0.89087566,0.23005985){\color[rgb]{0,0,0}\makebox(0,0)[lt]{\lineheight{1.25}\smash{\begin{tabular}[t]{l}$v$ [m]\end{tabular}}}}%
    \put(0.92046515,0.20405697){\color[rgb]{0,0,0}\makebox(0,0)[lt]{\lineheight{1.25}\smash{\begin{tabular}[t]{l}$4e-18$\end{tabular}}}}%
    \put(0.92046515,0.07314597){\color[rgb]{0,0,0}\makebox(0,0)[lt]{\lineheight{1.25}\smash{\begin{tabular}[t]{l}$-4e-18$\end{tabular}}}}%
    \put(0,0){\includegraphics[width=\unitlength,page=4]{./figures/shear_mod_nu049.pdf}}%
  \end{picture}%
\endgroup%

%% file: tikz/figure13.tex
\begin{tikzpicture}[scale=0.4, line cap=round]
	
		\scaling{0.4};
		
		
		\draw [thick] (5,5) rectangle (17,9);
		
		\normalsize
		
		\point{a}{17}{7};
		\point{b}{5}{7};
		\point{c}{11}{7};
		
		\support{2}{a};
		\support{2}{b};
		\support{2}{c}[90]; 
		
		\point{f}{5}{5};
		\point{g}{17}{5};

		\dimensioning{1}{f}{g}{3}[$a$];
		
		\draw [->, thick, bend left, >=latex] (3,5) to [edge label=$M$] (3,9);
		\draw [->, thick, bend right, >=latex] (19,5) to [edge label'=$M$] (19,9);
		
		\draw [->, thick, >=latex] (5,7) -- (7,7) node[right] {$x$};
		\draw [->, thick, >=latex] (5,7) -- (5, 10) node[above] {$y$};
		
		\draw [<->, thick, >=latex] (6,8.9) -- (6,7.1) node[midway, right] {$b$};
		\draw [<->, thick, >=latex] (6,6.9) -- (6,5.1) node[midway, right] {$b$};

	\end{tikzpicture}

%% file: tikz/figure16.tex
\begin{tikzpicture}[scale=0.5, line cap=round]
		
		\scaling{0.5};
		
		\draw [thick] (5,5) rectangle (17,9);
		
		\normalsize
		
		\point{a}{17}{5};
		\point{b}{17}{7};
		\point{c}{17}{9};
		
		\support{1}{a}[90];
		\support{1}{b}[90];
		\support{1}{c}[90];
		
		\point{d}{4}{8.75};
		\point{e}{4}{4.75};
		
		\lineload{4}{d}{e};
		\node at (3.25, 7) {$F$};
		
		\point{f}{5}{5};
		\point{g}{17}{5};
		
		\dimensioning{1}{f}{g}{3}[$a$];
		
		\draw [->, thick, >=latex] (5,7) -- (7,7) node[right] {$x$};
		\draw [->, thick, >=latex] (5,7) -- (5, 10) node[above] {$y$};
		
		\draw [<->, thick, >=latex] (6,8.9) -- (6,7.1) node[midway, right] {$b$};
		\draw [<->, thick, >=latex] (6,6.9) -- (6,5.1) node[midway, right] {$b$};
		
	\end{tikzpicture}